\documentclass{article}

\usepackage[english]{babel}

\usepackage[a4paper,top=2cm,bottom=2cm,left=3cm,right=3cm,marginparwidth=1.75cm]{geometry}

\usepackage{amssymb}
\usepackage{siunitx}
\PassOptionsToPackage{hyphens}{url}\usepackage{hyperref}
\usepackage{cleveref}
\usepackage[utf8]{inputenc}
\usepackage{csquotes}
\usepackage{booktabs}
\usepackage{longtable}
\usepackage{adjustbox}
\usepackage{array}
\usepackage{url}
\usepackage{titlesec}
\usepackage{authblk}
\usepackage{xcolor} % Load the xcolor package for color options

\newcommand{\bn}{\begin{equation}}
\newcommand{\en}{\end{equation}}
\newcommand{\ba}{\begin{array}{ll}}
\newcommand{\ea}{\end{array}}

\titleformat{\subsection}
  {\mdseries\itshape\large} % Medium series, italic shape, and large font size
  {\thesubsection}{1em}{} % Numbering, spacing, and the section title itself

\usepackage[english]{babel}
\usepackage{subcaption}

\crefformat{figure}{#2Figure~#1#3}
\Crefformat{figure}{#2Figure~#1#3}
\crefformat{table}{#2Table~#1#3}
\Crefformat{table}{#2Table~#1#3}
\crefformat{section}{#2Section~#1#3}
\Crefformat{section}{#2Section~#1#3}

\author{
    \large YiCheng Dai\footnote{\href{mailto:YichengDai@mail.ecust.edu.cn}{YichengDai@mail.ecust.edu.cn}}, 
    \large Wei Liao\footnote{\href{mailto:liaow@mail.ecust.edu.cn}{liaow@ecust.edu.cn}} \\
    \vspace{3mm}
    \large School of Physics, East China University of Science and Technology, \\
    130 Meilong Road, Shanghai 200237, P. R. China
}

\title{Constraint on Neutrino Statistics from Cosmological Data}

\begin{document}
\date{}
\maketitle

\begin{abstract}
We investigate the impact of neutrino statistical property on cosmology and the constraints imposed by cosmological data on neutrino statistics. 
Cosmological data from probes such as Cosmic Microwave Background(CMB) radiation and Baryon Acoustic Oscillation(BAO) are used to constrain the statistical parameter of neutrino. This constraint is closely related to the degeneracy effects among neutrino statistical property, the sum of neutrino masses, and the Hubble constant. Our results show that purely bosonic neutrinos can be ruled out at 95\% confidence level.
\end{abstract}

\section{Introduction}
Neutrinos are among the most elusive and least understood particles in the Standard Model(SM), primarily because they interact very weakly with matter and have extremely small masses. 
In recent years, precision experiments have improved constraints on the neutrino mass differences and hence limit the possible values for their sum.
For example, solar neutrino oscillation experiments can measure $\Delta m_{21}^2=m_2^2-m_1^2$, the mass squared difference of  neutrinos in mass eigenstates $\nu_1$ and $\nu_2$, and atmospheric neutrino oscillation experiments can measure $|\Delta m_{32}^2|=|m_3^2-m_2^2|$, the absolute value of
the mass squared difference of  neutrinos in mass eigenstates $\nu_2$ and $\nu_3$.
It can be inferred from  these measurements  that \cite{Esteban_2017, Gonzalez_Garcia_2021}
\bn
\begin{array}{ll}
\sum m_\nu\gtrsim 0.06\text{ eV}\quad(\mathrm{NH}),\\
\sum m_\nu\gtrsim 0.1\text{ eV}\quad(\mathrm{IH}),
\label{eq: oscillation}
\end{array}
\en
where NH refers to the case with $\Delta m_{32}^2>0$ and IH refers to the case  with $\Delta m_{32}^2<0$.
Cosmology also plays an important role in constraining the sum of neutrino masses. Observations of BAO and CMB are often combined to provide tighter constraints on cosmological models.  Since different datasets have different properties, the constraints on $\sum m_\nu$ would vary if different combination of datasets is chosen. For instance, the combination of Planck CMB data, WMAP 9-year CMB polarization data and a measurement of BAO from Baryon Oscillation Spectroscopic Survey, Sloan Digital Sky Survey, Wigglez LSS data and the 6dF galaxy redshift survey has provided a constraint $\sum m_\nu<0.23 \text{ eV}$ \cite{ABAZAJIAN201566,Planck2013}. When combining the Year 1 (Y1) data from the Dark Energy Spectroscopic Instrument (DESI) with CMB data from \textit{Planck} PR3 release and the Data Release 6 of Atacama Cosmology Telescope (ACT), an upper limit \cite{DESI:2024mwx}
\bn \sum m_\nu < 0.072\text{ eV} \quad (95\%), \en 
is achieved instead. Although this constraint seems more stringent, 
it may be in tension with limits derived from oscillation experiments {as shown in Eq.\ref{eq: oscillation}.}

Besides neutrino mass, the statistical nature of neutrinos is also of great interest. 
 Although neutrinos are generally believed to obey Fermi-Dirac(F-D)  statistics, there are very few experimental evidences supporting this hypothesis \cite{Dolgov_2005, Barabash_2007}. This makes studying the statistical property of neutrinos from a cosmological perspective important, because modifications to the neutrino statistics can have significant impacts on the evolution of the universe.
In some analysis \cite{deSalas:2018idd,CUCURULL_1996,Dolgov_2005_2,Dolgov_2005} , recent cosmological probes, including the CMB and Big Bang Nucleosynthesis (BBN), have been used to impose constraints on neutrino statistics. However, the bounds are still not stringent enough \cite{deSalas:2018idd}, and the possibility of  B-E statistics for neutrinos is allowed at  95\% confidence level. Therefore, it is essential to reconsider the statistical property of neutrinos from a cosmological perspective, in particular, to re-analyze the constraints on neutrinos using new data. 

In this paper, we extend the $\Lambda$CDM model to include neutrinos with  variable statistical property and fit the extended model to the latest cosmological data, primarily from \textit{Planck} PR3 and DESI. We find that the current data can exclude purely B-E neutrinos at the 95\% confidence level, while neutrinos with mixed statistics remain a viable option. The progress in constraining the statistical property of neutrinos is driven by the improved precision of the data and the enhanced capability to constrain cosmological parameters. Furthermore, our cosmological analysis using both CMB and BAO data indicates a preference for purely fermionic neutrinos. 

In the following sections, we first discuss the major physical effects of variable neutrino statistics and neutrino mass on the cosmological evolution . 
In \Cref{Data}, we briefly describe the cosmological datasets considered in this study. 
The fitting results of cosmological parameters are presented and discussed in \Cref{Results and Discussions}. 
Finally, in \Cref{Conclusions}, we summarize the main findings of  this study.

\section{Neutrino Statistics and Mass in Cosmology}
\label{Neutrino Statistics and mass in Cosmology}
To describe variable neutrino statistics,  we introduce a statistical parameter $\kappa_\nu$ which varies from $-1$ to $1$. 
For neutrinos in thermal equilibrium at temperature $T_\nu$, neutrinos have a distribution
\bn
f_\nu = \frac{1}{e^{E/T_\nu}+\kappa_\nu},
\label{eq:kappanu}
\en  
where $E$ is the energy of neutrino.  $\kappa_\nu=-1$ corresponds to B-E statistics, $\kappa_\nu=1$ corresponds to F-D statistics.
Using the distribution function, the energy density of thermal  neutrinos can be calculated as follows
\bn
\rho_\nu = g_\nu \int\frac{d^3 p}{(2\pi)^3}f_\nu(p) E_\nu(p) = g_\nu\int_0^\infty\frac{dp}{2\pi^2}\frac{p^2E(p)}{\mathrm{exp}( E(p)/T_\nu)+\kappa_\nu}
\en
where $g_\nu=6$ is the number of degrees of freedom of three flavors of neutrinos and anti-neutrinos.
If neutrinos obey F-D or B-E statistics, with $\kappa_\nu = 1$ or $-1$ respectively, 
the integral can be calculated analytically using Riemann $\zeta$ function in the relativistic limit, 
\bn
\rho_\nu = \left\{
\begin{array}{ll}
3\times7/8\left(T_\nu/T\right)^4\times \rho_\gamma \quad \text{for $\kappa_\nu = 1$} \\
3\times\left(T_\nu/T\right)^4\times \rho_\gamma \quad  \text{for $\kappa_\nu = -1$} 
\end{array}
\right.,
\en
where $T$ and $\rho_\gamma$ are, respectively, the temperature and energy density of photons. 
In the standard model of cosmic evolution,  relativistic neutrinos decouple from the thermal bath earlier than the annihilation of electrons and positrons to photons.
The entropy of photons is increased, and, it can be shown that $T_\nu/ T\approx (4/11)^{1/3}$. 
For models of variable neutrino statistics, this result is still valid.
{Therefore, in the relativistic limit, the energy density of B-E neutrino is larger than that of the F-D neutrino by a factor of 8/7. }

{
In non-relativitic regime, $\rho_\nu$ can be approximately written as $\rho_\nu\approx m\times n_\nu$, where $m$ is the  mass of neutrino and $n_\nu$ is the number density. 
The number density at scale factor $a$ is related to the number density at decoupling $n_\nu(a_\mathrm{dec})$ by
\bn
n_\nu(a) a^3 = n_{\nu}(a_\mathrm{dec}) a_\mathrm{dec}^3.
\en
Equipped with this we can calculate 
\bn
\rho_\nu(a) = m_\nu  \frac{a_\mathrm{dec}^3}{a^3}  n_{\nu}(a_\mathrm{dec}),
\en
where
\bn
\begin{array}{ll}
n_{\nu}(a_\mathrm{dec}) &= 
  \displaystyle
  g_\nu \int_0^\infty \frac{dp}{2\pi^2}\,
  \frac{p}{\exp\bigl(E(p)/T_{\nu\mathrm{dec}}\bigr) + \kappa_\nu}
  \\
&=
\displaystyle
  \left\{
  \begin{array}{ll}
  \displaystyle
  g_\nu\,T_{\nu\mathrm{dec}}^3/(2\pi^2)\times\zeta(3)\times\Gamma(3)\times\frac{3}{4},
  & \text{for } \kappa_\nu = 1,\\[6pt]
  \displaystyle
  g_\nu\,T_{\nu\mathrm{dec}}^3/(2\pi^2)\times\zeta(3)\times\Gamma(3),
  & \text{for } \kappa_\nu = -1.
  \end{array}
  \right.
\end{array}
.
\en
Therefore, in the non-relativistic limit, the energy density of B-E neutrino is larger than that of the F-D neutrino by a factor of 4/3.

Taken together, these factors (8/7 when relativistic and 4/3 when non-relativistic) show how neutrino statistics significantly affect their energy density during the evolution.}
 
The impact of neutrinos, especially massless ones, on the CMB can largely be understood by their effects on the sound horizon \cite{Hou_2013, Lesgourgues}.
The sound horizon can be defined as:
\bn
r_s(\eta) = \int_{t_\mathrm{in}}^t\frac{c_s(t)dt}{a(t)} = \int_0^a\frac{c_sda}{a^2H},
\label{eq:rseta}
\en
where the sound speed
\bn c_s^2(\eta)\equiv\frac{1}{3\left(1+R\right)}\en
is determined by the baryon and photon density, $\rho_b$ and $\rho_\gamma$, through $R\equiv 3\rho_b/(4\rho_\gamma)$. 
It is known that the Hubble expansion rate
\bn
H = H_0\sqrt{\Omega_r a^{-4} + \Omega_m a^{-3} + \Omega_\Lambda}\propto
\left\{
\begin{array}{ll}
a^{-2}\quad\text{In radiation domination}\\
a^{-3/2}\quad \text{In matter domination}
\end{array}
\right.
\label{eq:Hubble}
\en
is larger during the radiation-dominated epoch compared with during the matter-dominated epoch.
If the equality time is delayed, the sound horizon at recombination will be smaller.

For massless neutrinos, a smaller $\kappa_\nu$ would increase the density of relativistic species.
This would both delay the equality time and increase $H(a)$ in radiation-dominated epoch.
The combined effects would make the sound horizon at recombination smaller.
For massive neutrinos, $\kappa_\nu$ would affect both the radiation energy density in the early time
and the matter energy density in the late time. So a more detailed discussion is needed in this case.

\begin{figure}[htbp]
\centering
\begin{subfigure}[b]{0.49\textwidth}
\centering
\includegraphics[width=\textwidth]{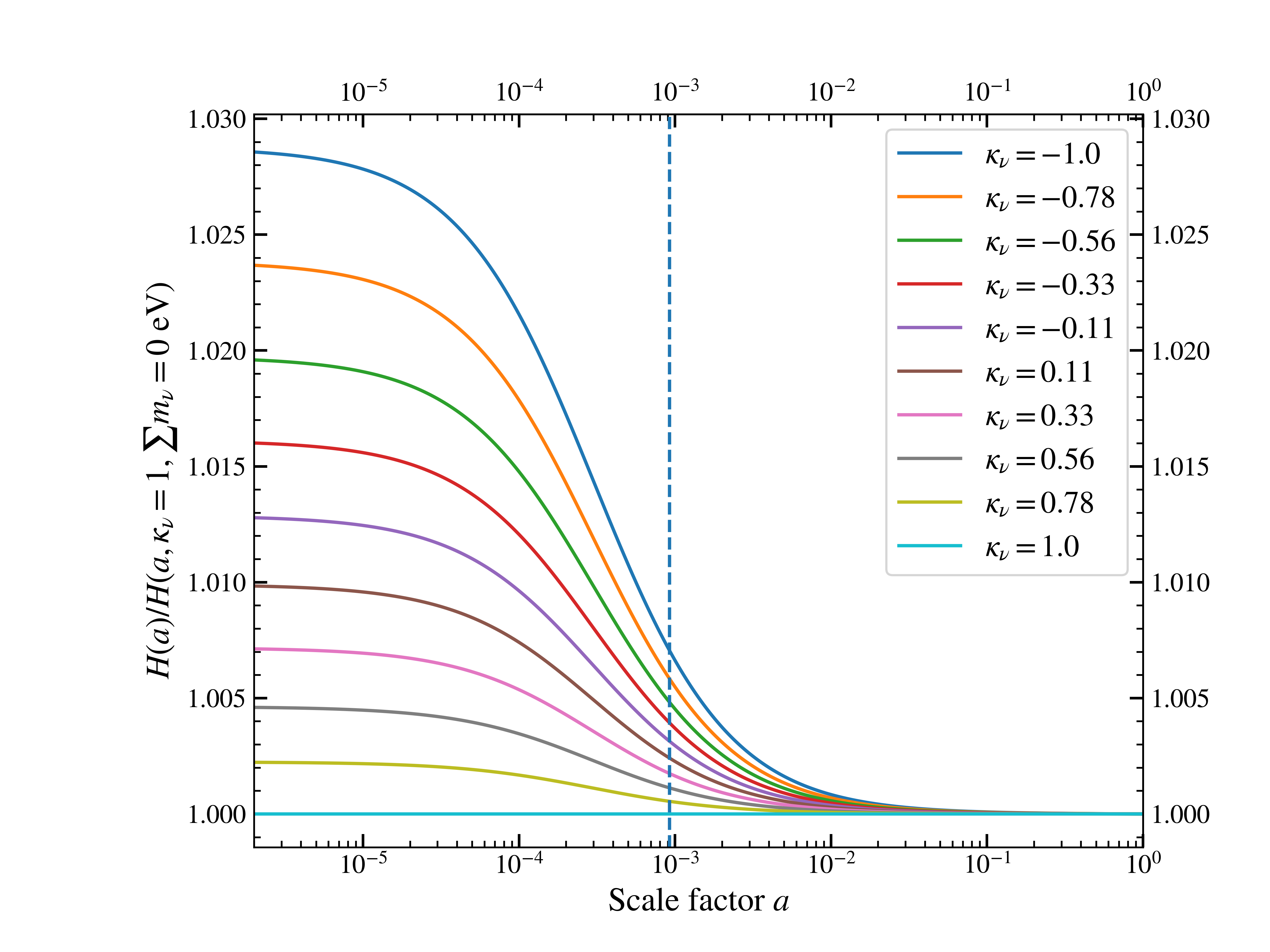}
\caption{}
\label{fig:Hubble_normalized_kappa_vary_sub1}
\end{subfigure}
\hfill
\begin{subfigure}[b]{0.49\textwidth}
\centering
\includegraphics[width=\textwidth]{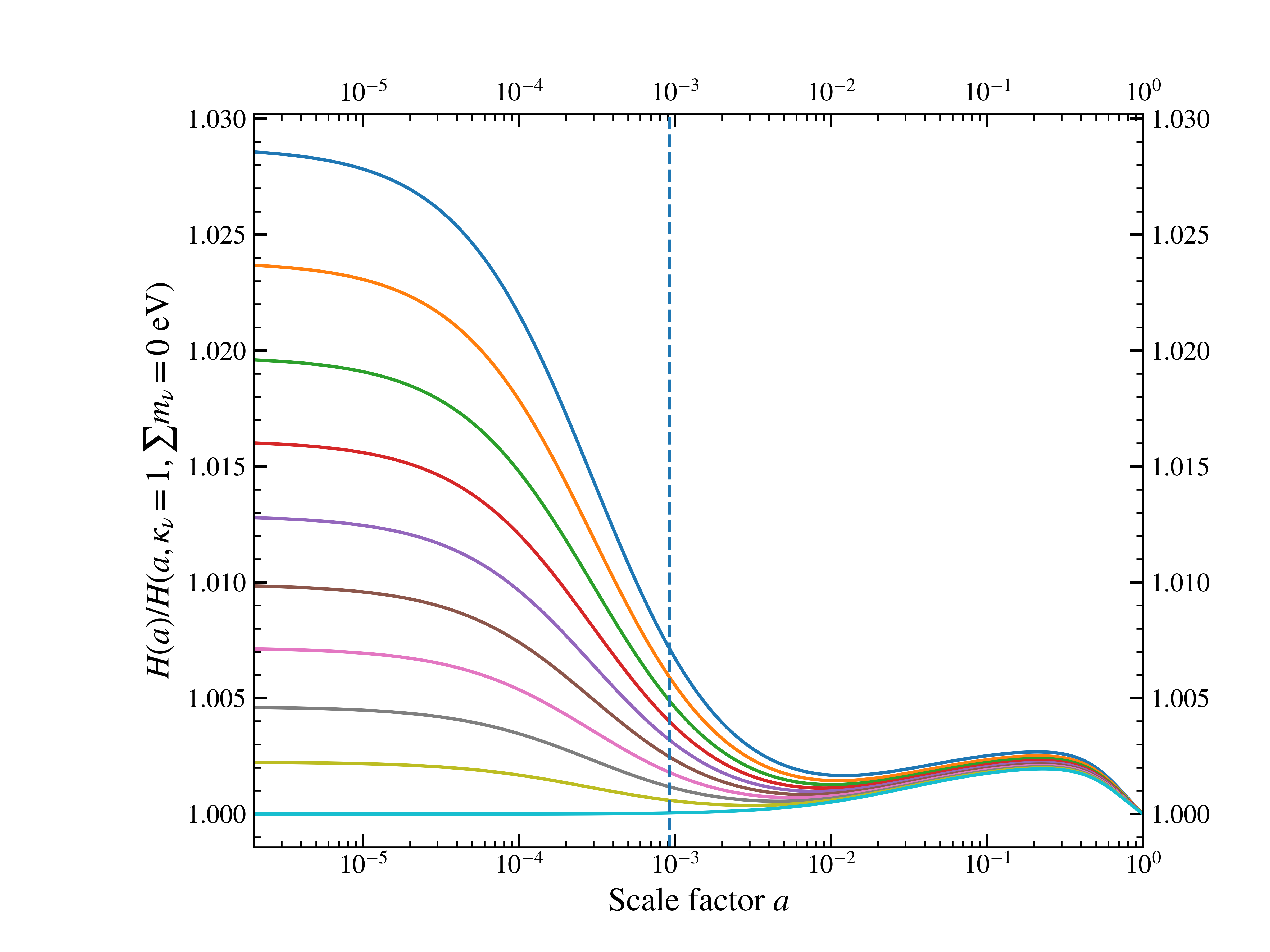}
\caption{}
\label{fig:Hubble_normalized_kappa_vary_sub2}
\end{subfigure}
\begin{subfigure}[b]{0.48\textwidth}
\centering
\includegraphics[width=\textwidth]{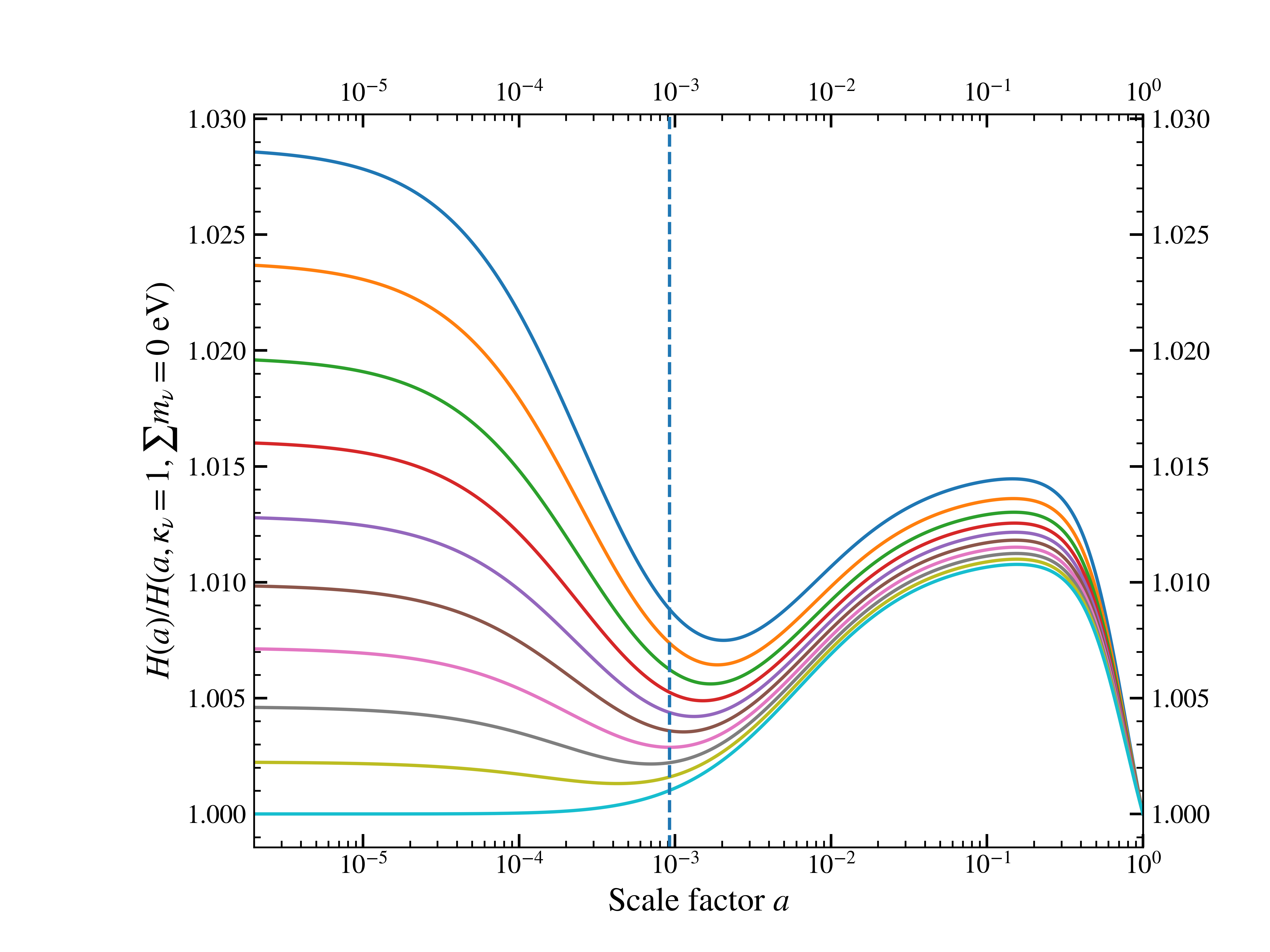}
\caption{}
\label{fig:Hubble_normalized_kappa_vary_sub3}
\end{subfigure}
\begin{subfigure}[b]{0.48\textwidth}
\centering
\includegraphics[width=\textwidth]{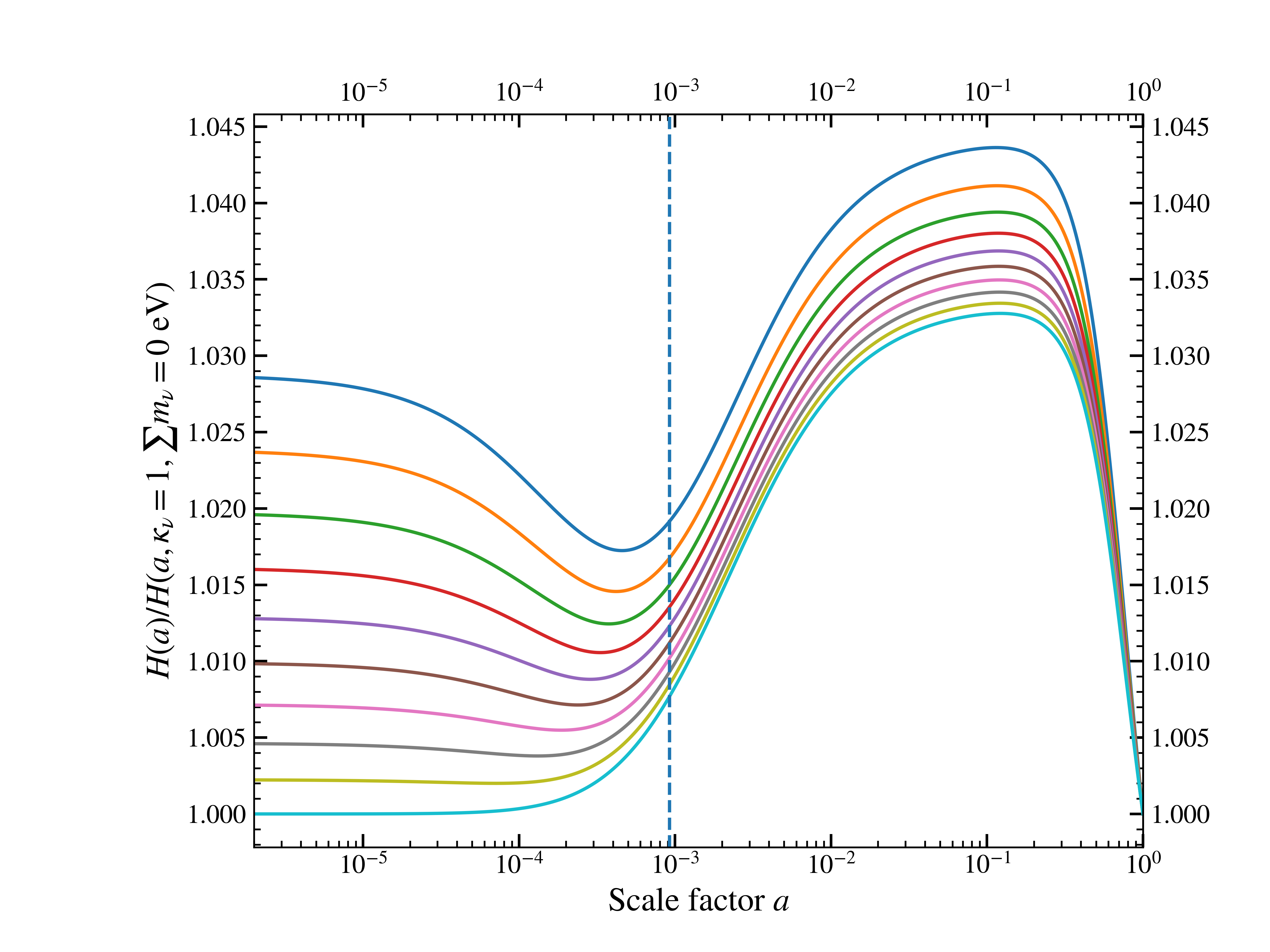}
\caption{}
\label{fig:Hubble_normalized_kappa_vary_sub4}
\end{subfigure}
\caption{Evolution of Hubble expansion rate $H(a)$ with the scale factor $a$ for different statistical parameter $\kappa_\nu$. Each curve is normalized to the case of $\kappa_\nu = 1$  and $\sum m_\nu=0.0$eV.  For (a), $\sum m_\nu = 0$ eV and $H_0=67$ $\mathrm{km}$~$\mathrm{s}^{-1} \mathrm{Mpc}^{-1}$ are held fixed; for (b),  $\sum m_\nu = 0.06$ eV and $H_0=67$ $\mathrm{km}$~$\mathrm{s}^{-1} \mathrm{Mpc}^{-1}$; for (c), $\sum m_\nu = 0.3$ eV and $H_0=67$ $\mathrm{km}$~$\mathrm{s}^{-1} \mathrm{Mpc}^{-1}$; for(d), $\sum m_\nu = 0.9$ eV and $H_0=67$ $\mathrm{km}$~$\mathrm{s}^{-1} \mathrm{Mpc}^{-1}$. The CMB temperature at present $T_\mathrm{CMB} = 2.73$~K, baryon energy density $\omega_\mathrm{b}\equiv \Omega_\mathrm{b}h^2 = 0.0224$ 
and cold dark matter density $\omega_\mathrm{cdm}\equiv \Omega_\mathrm{cdm}h^2=0.1201$ are also fixed in this and the following three plots in this section. The vertical dashed line in each subplot indicates the scale factor at recombination, $a_\mathrm{rec}$, for the case of $\kappa_\nu = 1$.
}
\label{fig:Hubble_normalized_kappa_vary}
\end{figure}

To illustrate the effect of neutrino mass in cosmological evolution, we numerically compute the
Hubble expansion rate and give in Fig.~\ref{fig:Hubble_normalized_kappa_vary} the evolution of $H(a)$ for different $\kappa_\nu$
where in (a): $\sum m_\nu=0$ eV and $H_0=67$ $\mathrm{km}$~$\mathrm{s}^{-1} \mathrm{Mpc}^{-1}$ are held fixed;
 (b): $\sum m_\nu= 0.06$ eV and $H_0=67$ $\mathrm{km}$~$\mathrm{s}^{-1} \mathrm{Mpc}^{-1}$;
 (c) $\sum m_\nu=0.3$ eV and $H_0=67$ $\mathrm{km}$~$\mathrm{s}^{-1} \mathrm{Mpc}^{-1}$;
 (d) $\sum m_\nu=0.9$ eV and $H_0=67$ $\mathrm{km}$~$\mathrm{s}^{-1} \mathrm{Mpc}^{-1}$.
In this figure and the remaining three figures in this section, we fix CMB temperature  at present $T_\mathrm{CMB} = 2.73$ K, 
baryon energy density $\omega_\mathrm{b}\equiv \Omega_\mathrm{b}h^2 = 0.0224$ 
and cold dark matter density $\omega_\mathrm{cdm}\equiv \Omega_\mathrm{cdm}h^2 = 0.1201$,
and we assume the universe is spatially flat.
We can see in the figure that smaller $\kappa_\nu$ always gives larger $H(a)$ irrespective of the chosen value of $\sum m_\nu$ 
in both radiation-dominated and matter-dominated eras, as expected.  
This means that smaller $\kappa_\nu$ should always give smaller sound horizon, according to Eq.(\ref{eq:rseta}).
The differences between curves of $\sum m_\nu=0$ eV and curves of $\sum m_\nu=0.06$ eV are actually very small.
One can notice that all the lines approach $1$  as the scale factor $a$ appoaches 1. This is because
we have used a fixed $H_0$ in these plots which means that the total energy density at the present time has been chosen as a fixed value.
{One can also notice that the effects of neutrino statistics is more significant in the early radiation-dominated era than in the late matter-dominated era.}
{Although the variation remains at the level of a few percent even in the radiation-dominated era, it may have important implications in high-precision cosmological data analyses.}

\begin{figure}[htbp]
\centering
\begin{subfigure}[b]{0.49\textwidth}
\centering
\includegraphics[width=\textwidth]{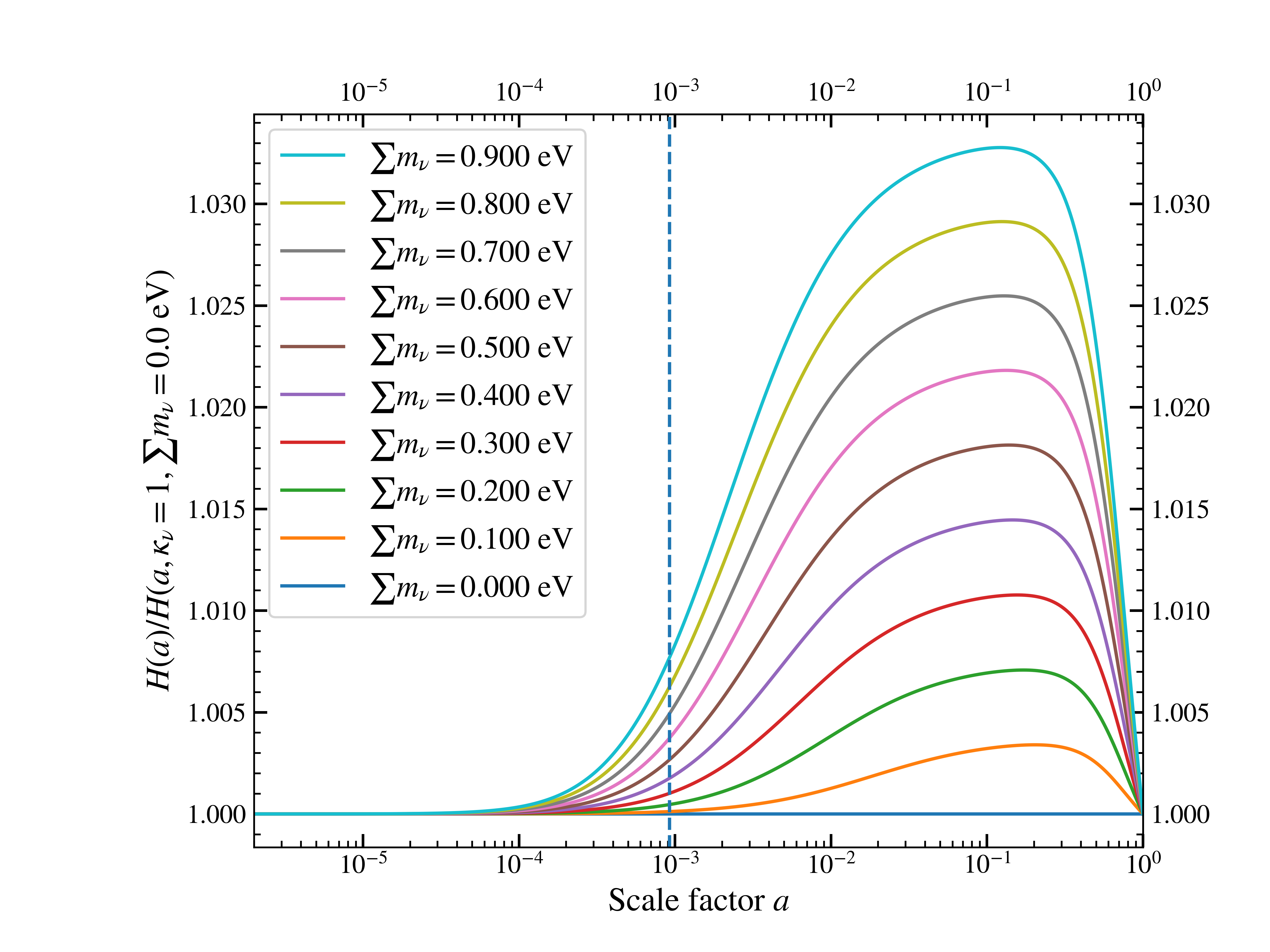}
\caption{}
\label{fig:Hubble_normalized_mass_vary_sub1}
\end{subfigure}
\hfill
\begin{subfigure}[b]{0.49\textwidth}
\centering
\includegraphics[width=\textwidth]{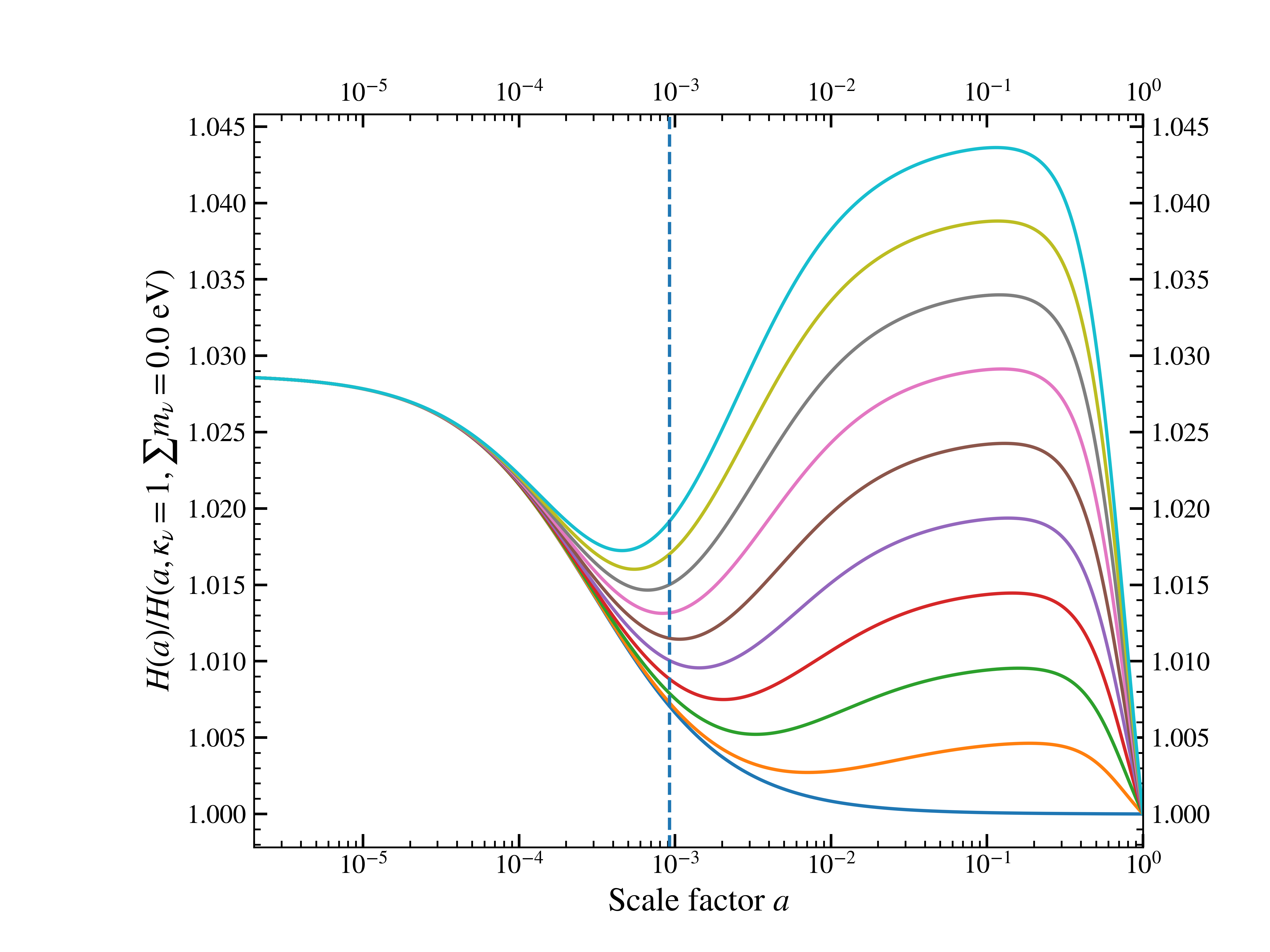}
\caption{}
\label{fig:Hubble_normalized_mass_vary_sub2}
\end{subfigure}
\caption{Evolution of Hubble expansion rate $H(a)$ with the scale factor $a$ for different $\sum m_\nu$. All curves are normalized to the case of $\kappa_\nu = 1$  and $\sum m_\nu=0.0$ eV. For (a), $\kappa_\nu = 1$ and $H_0=67$ $\mathrm{km}$~$\mathrm{s}^{-1} \mathrm{Mpc}^{-1}$ are held fixed; for (b), $\kappa_\nu = -1$  and $H_0=67$ $\mathrm{km}$~$\mathrm{s}^{-1} \mathrm{Mpc}^{-1}$. The vertical dashed line in each subplot indicates the scale factor at recombination, $a_\mathrm{rec}$, for the case of $\sum m_\nu = 0$ eV.}
\label{fig:Hubble_normalized_mass_vary}
\end{figure}

Massive neutrinos become non-relativistic in matter-dominated era, contributing to matter energy density $\Omega_m$. 
This results in a significant increase in $H(a)$ in late times.
In Fig.~\ref{fig:Hubble_normalized_mass_vary}, we illustrate the effect of neutrino mass on the evolution of Hubble expansion rate $H(a)$ with
(a) $\kappa_\nu = 1$  and (b) $\kappa_\nu = -1$. 
 One can find  that a larger $\sum m_\nu$ leads to a larger $H(a)$ in the matter-dominated era
and in particular near the epoch of recombination. 
This means that a larger neutrino mass should also lead to a smaller sound horizon, according to Eq.(\ref{eq:rseta}).
Notice that this effect of neutrino mass is more significant in the matter-dominated era than in the radiation-dominated era.
Since the contribution to the sound horizon is determined by the contribution before recombination,
the effect of neutrino statistics on the $r_s$ is more significant than the effect of neutrino mass. 
Furthermore, if $\sum m_\nu$ is significantly small, for example $\sum m_\nu\lesssim0.1$ eV, neutrino mass would have a very mild effect on the evolution of the universe as shown by the curves for $\sum m_\nu = 0.1$ eV in Fig.~\ref{fig:Hubble_normalized_mass_vary}.
{This effect is roughly at the level of 0.1\%, which is much smaller than the effect induced by $\kappa_\nu$ during the radiation-dominated era in Fig.~\ref{fig:Hubble_normalized_kappa_vary}.}

\begin{figure}[htbp]
\centering
\begin{subfigure}[b]{0.48\textwidth}
\centering
\includegraphics[width=\textwidth]{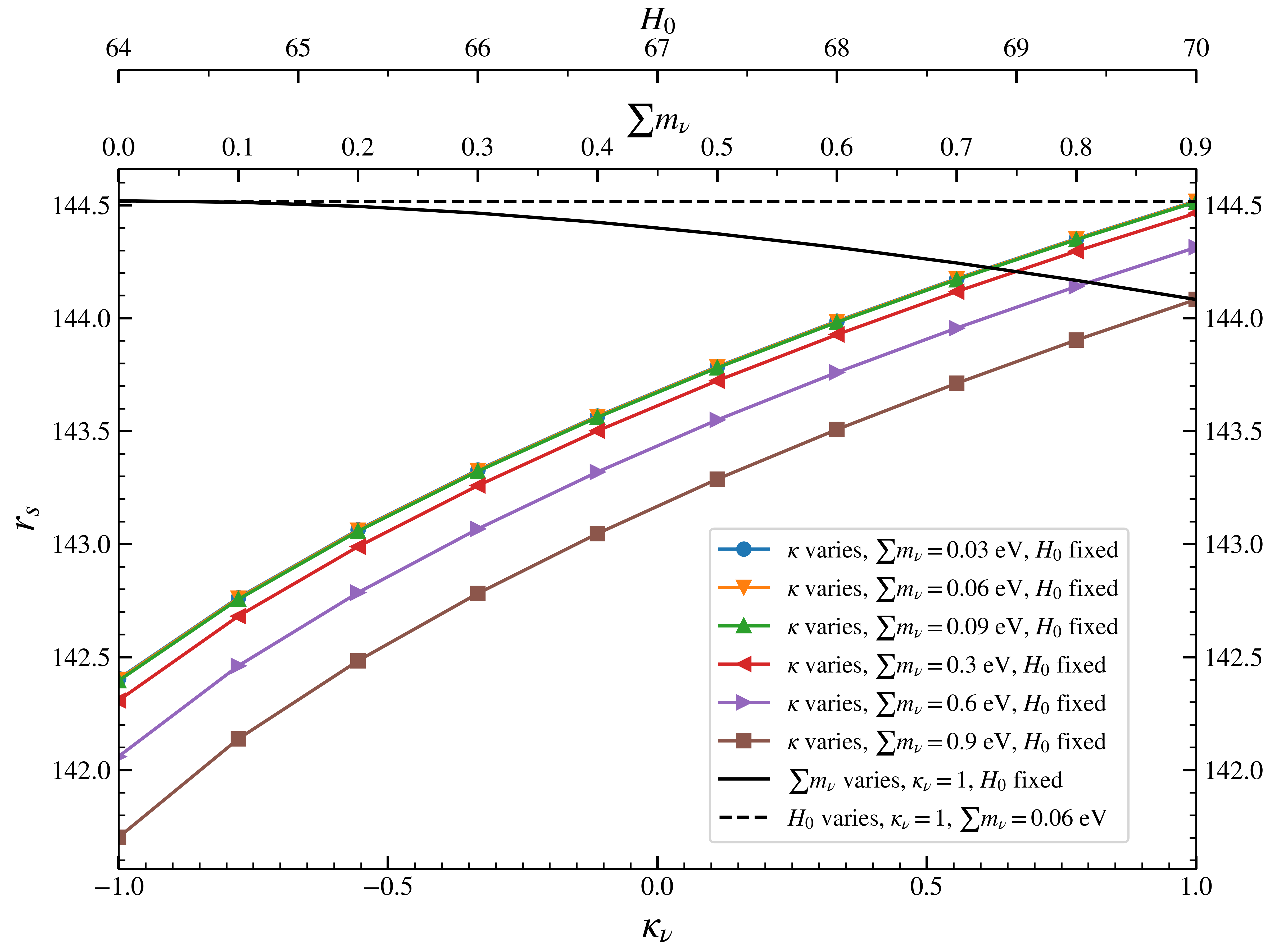}
\caption{}
\label{fig:rs_da_H0_theta_sub1}
\end{subfigure}
\hfill
\begin{subfigure}[b]{0.48\textwidth}
\centering
\includegraphics[width=\textwidth]{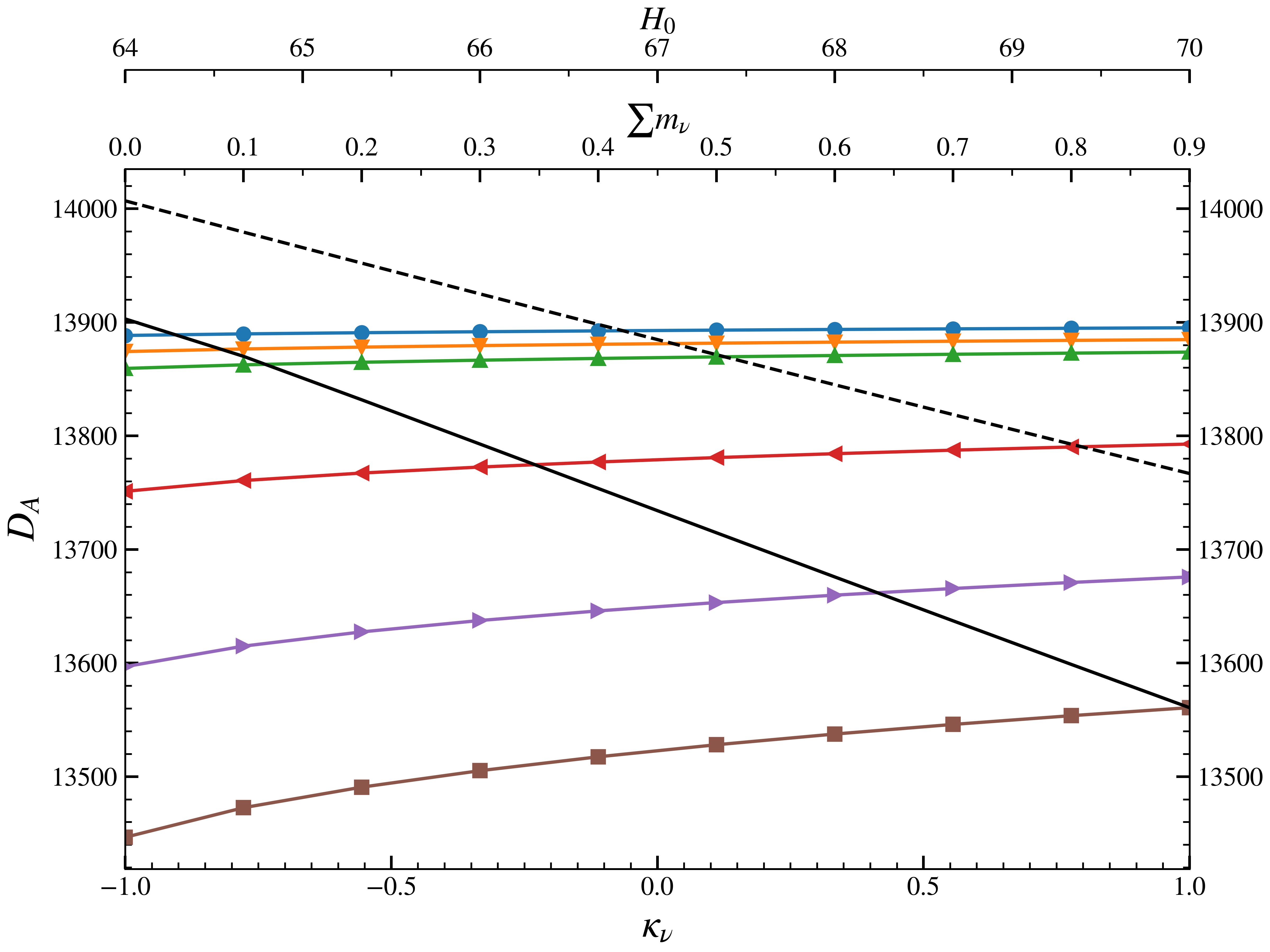}
\caption{}
\label{fig:rs_da_H0_theta_sub2}
\end{subfigure}
%\vspace{15mm}
\begin{subfigure}[b]{0.48\textwidth}
\centering
\includegraphics[width=\textwidth]{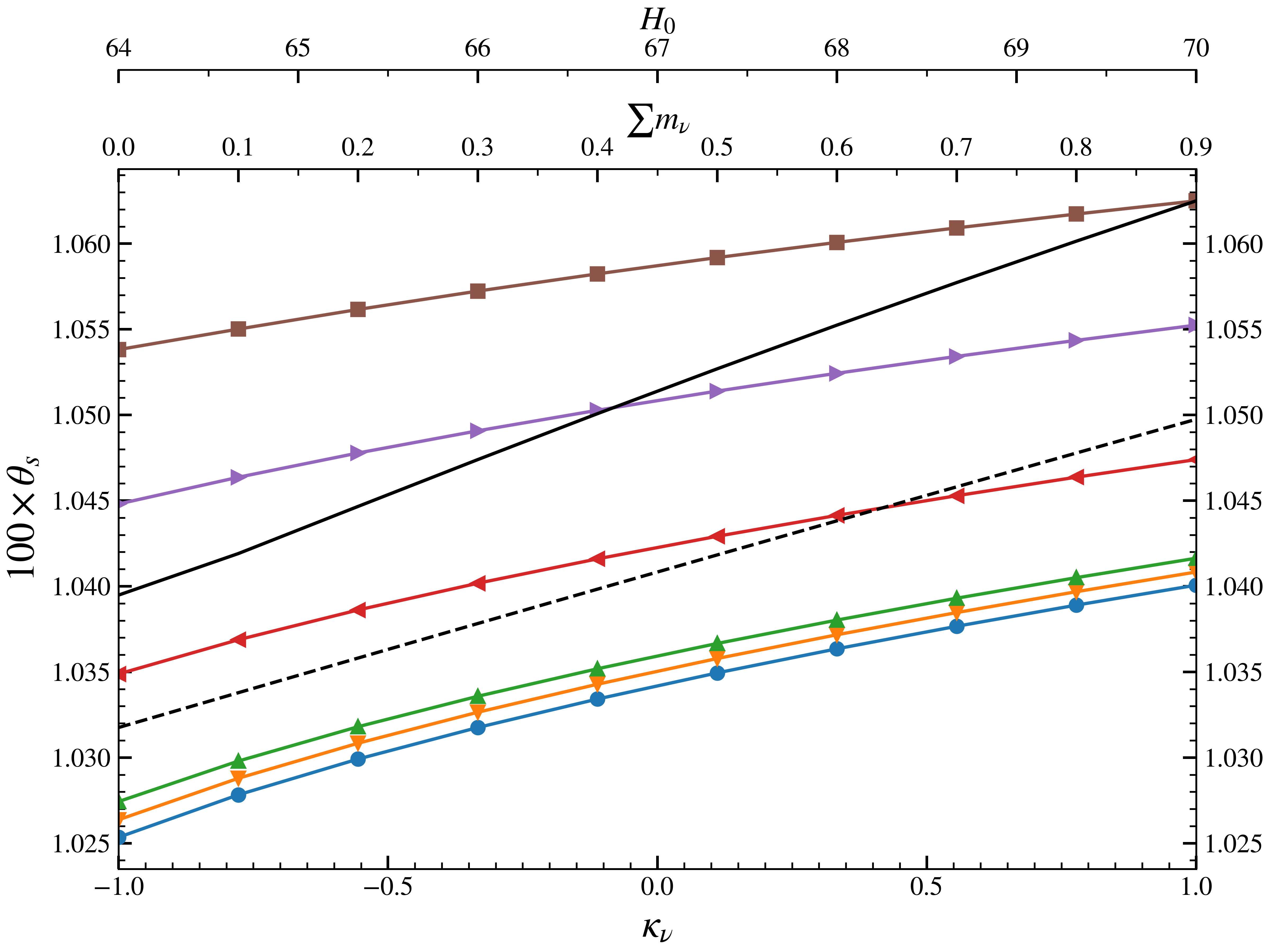}
\caption{}
\label{fig:rs_da_H0_theta_sub3}
\end{subfigure}
\caption{The sound horizon $r_s(\eta)$, angular diameter $D_A(\eta)$ and peak scale parameter $\theta_s(\eta)$ at recombination.
The variation with respect to $\kappa_\nu$ are illustrated by the solid labeled (colored) lines in the subplots, where each label (color) corresponds to a different $\sum m_\nu$. The variations with respect to $\sum m_\nu$ are depicted by solid unlabeled (black) lines.
The dashed line describes the variations caused by $H_0$.
All the numerical results are calculated with a modification of the Boltzmann code {CLASS} \cite{DiegoBlas2011}.}
\label{fig:rs_da_H0_theta}
\end{figure}

The above mentioned effects on the sound horizon are illustrated in Fig.~\ref{fig:rs_da_H0_theta}(a). 
The colored and labeled lines are plots for $r_s$ versus $\kappa_\nu$ with $H_0$ fixed to 67 $\mathrm{km}$ $\mathrm{s}^{-1} \mathrm{Mpc}^{-1}$.
One can see that smaller $\kappa_\nu$ always gives smaller sound horizon irrespective of the chosen value of $\sum m_\nu$.
The effect of neutrino mass can also be seen clearly in the solid and unlabelled line in Fig.~\ref{fig:rs_da_H0_theta}(a), 
for which $\kappa_\nu$ is fixed to 1 and $\sum m_\nu$ varies from 0 eV to 0.9 eV.
We can see that larger neutrino mass gives smaller $r_s$, in agreement with the expectation discussed above.
However, this effect is not as significant as varying the statistics of neutrino, as can be seen in the figure, which is also in agreement with the discussions above.

The statistical parameter $\kappa_\nu$ also affects the angular diameter at recombination
\bn
D_A(z)\propto\int_0^z\frac{dz^\prime}{H(z^\prime)},
\en
where the integrand is proportional to the inverse of Hubble rate $H(a)$. 
 Major contribution to $D_A(\eta)$ primarily comes from the integration in the late time regime, which is different from the case of $r_s$.
As discussed for Fig.~\ref{fig:Hubble_normalized_kappa_vary},  larger $\kappa_\nu$ gives smaller Hubble expansion rate $H(a)$  throughout the entire evolution. 
So, larger $\kappa_\nu$ should also give rise to larger $D_A$, as illustrated by the solid labeled lines in Fig.~\ref{fig:rs_da_H0_theta}(b).
However, the effect of neutrino statistics is more significant in the radiation-dominated era than in the matter-dominated era, as discussed above, 
so the effect of varying $\kappa_\nu$ on $D_A(\eta)$ is less significant than on $r_s$. 
This effect can be seen clearly  in solid labeled lines in Fig.~\ref{fig:rs_da_H0_theta}(b).
Similarly, larger $\sum m_\nu$ should lead to smaller $D_A$, as can be seen in solid unlablled line in Fig.~\ref{fig:rs_da_H0_theta}(b).
We further note that the effect of neutrino mass is more significant in the late matter-dominated era than in the radiation-dominated era, 
so the effect of varying $\sum m_\nu$ is much more significant on $D_A$ than on $r_s$.
This point can be seen clearly in Fig.~\ref{fig:rs_da_H0_theta}(b). 

Finally,  we can study the combined effects of both $\kappa_\nu$ and $\sum m_\nu$ on peak scale parameter 
\bn
\theta_s \equiv r_s/D_A,
\en
which is an approximation to the peak location of CMB angular power spectrum.
In Fig.~\ref{fig:rs_da_H0_theta}(c),  $100\times \theta_s$ versus $k_\nu$ and $100\times \theta_s$ versus  $\sum m_\nu$
are plotted with solid labled lines and solid unlabeled line respectively. Parameters chosen for each line are the same for corresponding line 
in Fig.~\ref{fig:rs_da_H0_theta}(a).
We note that increasing $\kappa_\nu$ increases both $r_s$ and $D_A$, but the effect on $r_s$ is more significant.
So the $\theta_s$ increases with increasing $\kappa_\nu$, as can be seen in the solid lablled lines in Fig.~\ref{fig:rs_da_H0_theta}(c).
On the other hand, increasing $\sum m_\nu$ decreases both $r_s$ and $D_A$, but the effect on $D_A$ is more significant.
So $\theta_s$ also increases with increasing $\sum m_\nu$, as can be seen in the solid unlablled lines in Fig.~\ref{fig:rs_da_H0_theta}(c).

We further note that, when $\sum m_\nu$ is particularly small, specifically $\sum m_\nu\lesssim0.1$ eV, neutrinos remain relativistic until very late times. 
Such small neutrino masses are always favored in cosmological analyses and have little impact on $H(a)$. 
Consequently, the variation of $r_s$, $D_A$ and $\theta_s$ caused by $\sum m_\nu$ in this small mass region  are very small.
As can be seen in Fig.~\ref{fig:rs_da_H0_theta}(c), three solid labelled lines corresponding to $\sum m_\nu = 0.03, 0.06, 0.09$ eV
are very close to each other, which is in agreement with our expectation.

\begin{figure}[htbp]
\centering
\includegraphics[width=0.48\textwidth]{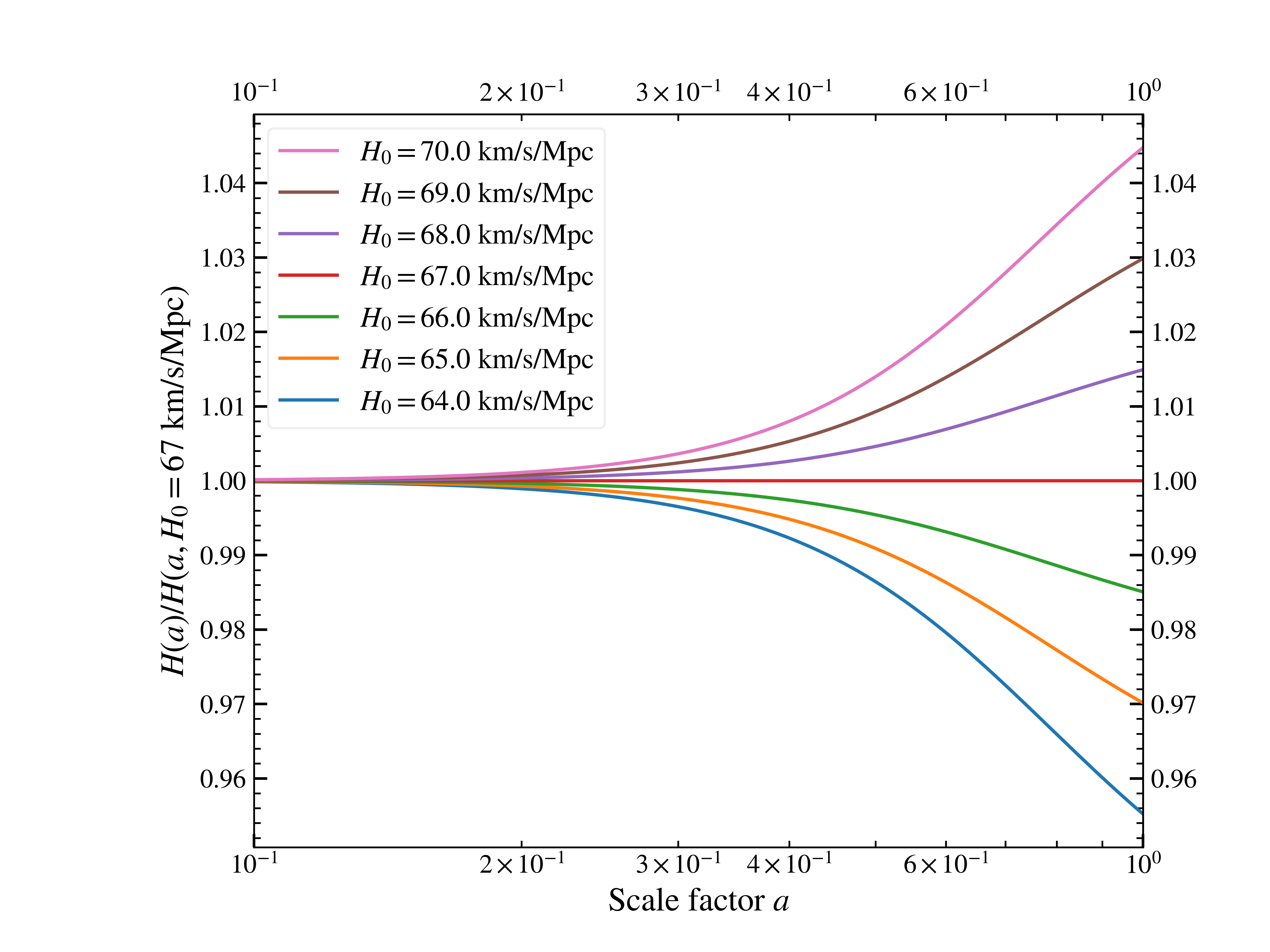}
\caption{Evolution of Hubble expansion rate $H(a)$ with the scale factor $a$ for varying the present Hubble expansion rate $H_0$. 
{All curves are normalized to the corresponding one with $H_0 = 67$~$\mathrm{km}$~$\mathrm{s}^{-1}~\mathrm{Mpc}^{-1}$, which falls within the $1\sigma$ range of the \textit{Planck} best-fit estimate, $H_0 = (67.4 \pm 0.5)$~$\mathrm{km}$~$\mathrm{s}^{-1}~\mathrm{Mpc}^{-1}$ \cite{Planck2018results}.} 
$\kappa_\nu = 1$ and $\sum m_\nu = 0.06$ eV are held fixed. }
\label{fig:Hubble_normalized_H_0_vary}
\end{figure}

Besides $\kappa_\nu$ and $\sum m_\nu$, the effects of $H_0$ are also noteworthy. 
$H_0$ is the present Hubble expansion rate which is determined by the present energy density of the universe.
Changing $H_0$ does not have significant effect on the expansion rate in the early time, in particular in the radiation-dominated era,
if we fix $T_\mathrm{CMB}$. 
It certainly has important effect on $H(a)$ in the late universe.
In Fig.~\ref{fig:Hubble_normalized_H_0_vary} , we plot $H(a)$ versus $a$ for a set of $H_0$. 
We can see clearly that $H_0$ mainly affects the late time $H(a)$.
So, $r_s(\eta)$ is basically unchanged for varying $H_0$, as can be seen  in the dashed line in Fig.~\ref{fig:rs_da_H0_theta}(a),
and $D_A(\eta)$ decreases as $H_0$ increases, as can be seen  in the dashed line in Fig.~\ref{fig:rs_da_H0_theta}(b).
As a result, $\theta_s$  increases with $H_0$, as can be seen  in dashed line in Fig.~\ref{fig:rs_da_H0_theta}(c).

Referring to Fig.~\ref{fig:rs_da_H0_theta}(c), 
we find that increasing any of the three parameters $\kappa_\nu$, $\sum m_\nu$ or $H_0$, would have similar influences on $\theta_s(\eta)$. 
If $\theta_s(\eta)$ at recombination is a known value from experiments, 
we can expect that there should be a degeneracy among $\kappa_\nu$, $\sum m_\nu$ and $H_0$ in the final results of fitting. 
Besides, if the sum of neutrino mass $\sum m_\nu$ is significantly small, 
it would be less likely to have a significant impact on the evolution of the late universe 
and the correlation between $\kappa_\nu$ and $H_0$ would be enhanced.

\section{Tools, Datasets and Models}
\label{Data}
{As mentioned in previous sections, introducing different datasets into the fitting analysis would cause various constraints. 
In the past few years, many cosmological observations, like \textit{Planck} and DESI, have released their latest data with higher accuracy. 
 Although neutrino statistics have already been analysed before \cite{deSalas:2018idd},  only very weak bounds are obtained on neutrino statistics. 
So, it is worthwhile to introduce new datasets into fitting and check the neutrino statistics in more detail.

In this section, we briefly describe our analysis prescription, along with the datasets and codes used.}

Our study is based on the $\Lambda$CDM model extended by two parameters:  neutrino statistics parameter $\kappa_\nu$ defined above and degenerate neutrino masses $\sum m_\nu$. Initially, we fit the model using {CMB} data and {CMB+BAO} data respectively to assess whether the latest datasets can impose  more stringent constraints on $\kappa_\nu$.
We refer to the fitting using CMB data as \textbf{CMB} and the fitting using {CMB+BAO} data as \textbf{CMB+BAO} in later discussions.
The details for their data combinations are provided below.
To further investigate the degeneracy between $H_0$ and $\kappa_\nu$, we fit the model using the {CMB+BAO+SNe} dataset which includes the Pantheon supernovae sample and a Gaussian constraint on the supernova absolute magnitude from \cite{Riess2021}.
We refer to the fitting using this dataset as \textbf{CMB+BAO+SNe}.
{Early-Universe probes, such as CMB typically yield a lower Hubble constant of $H_0 = (67.4 \pm 0.5)$~$\mathrm{km}$ $\mathrm{s}^{-1} $~$\mathrm{Mpc}^{-1}$ \cite{Planck2018results, louis2025atacamacosmologytelescopedr6}.
By contrast, local measurements favor a higher value near $H_0 = (73.0 \pm 1.0)$~$\mathrm{km}$~$\mathrm{s}^{-1}$~$\mathrm{Mpc}^{-1}$ \cite{Riess_2011,Riess_2016,Riess_2022}.
This discrepancy is often referred to as the 'Hubble tension' \cite{kamionkowski2022hubbletensionearlydark}.
Therefore the resulting constraints should be interpreted with caution, as the combined dataset exhibits significant internal tension.

For the {CMB} data, we combine the baseline temperature (TT) and polarization (EE) auto-spectra, along with their cross-spectra (TE), as incorporated in the {Commander} likelihood for $\ell < 30$ and the {plik} likelihood for $\ell > 30$ from the PR3 release \cite{Planck2018results}.
We further combine the CMB lensing data, which consists of the NPIPE PR4 Planck CMB lensing reconstruction and Data Release 6 from the Atacama Cosmology Telescope (ACT) \cite{ACT_Madhavacheril_2024,ACT_Qu_2024,ACT_Carron_2022}.
The mean value of $H_0$ given by \textit{Planck} PR3 is not significantly higher than the one given by \textit{Planck} PR4.
However, the constraint has become tighter.

For {BAO} dataset, DESI BAO distance measurements obtained from galaxies, quasars and Lyman-$\alpha$ tracers are included \cite{desicollaboration2024desi2024iiibaryon,desicollaboration2024desi2024ivbaryon}. The inclusion of {BAO} data prefers a lower or even negative neutrino mass \cite{craig2024nusgoodnews}. 

To further study the effects  of degeneracy between $H_0$ and $\kappa_\nu$, local measurement on $H_0$ is included for \textbf{CMB+BAO+Sne}. For a better precision, Pantheon analysis of Type Ia supernovae is also imposed to constrain cosmological parameters \cite{Riess2021,Brout2022}. 
Due to the significant Hubble tension, this dataset combination cannot be regarded as statistically well-founded, and its resulting constraints should be interpreted with particular caution.
However, it could still provide comparative insights for interpreting other results.

The theoretical models are computed with a modification of  the Boltzmann code {CLASS} \cite{DiegoBlas2011}. All Bayesian inference is performed by {Cobaya} \cite{Cobaya1,Cobaya2}, using Metropolis-Hastings MCMC sampler \cite{MCMC1,MCMC2}.
{We put flat priors on parameters listed in Table~\ref{tab:priors}}.
For each dataset and model combination, we take advantage of oversampling by setting the parameter {oversample\_power} to 0.4, using the dragging method \cite{MCMC3} and running four chains in parallel. If Gelman-Rubin criterion \cite{Gelman-Rubin} $R-1<0.01$ is satisfied, chains will be stopped. For plotting the posteriors, we use {getdist} \cite{lewis2019getdistpythonpackageanalysing}.

\begin{table}[htbp]
    \centering
    \label{tab:cosmo_params}
{   
    \begin{tabular}{l l l l}
    \toprule
    \textbf{Parameter} & \textbf{Prior range} & \textbf{Definition} \\
    \midrule
    
    $\Omega_\mathrm{b} h^2$     	& [0.005, 0.1]                   & Baryon density parameter \\
    $\Omega_\mathrm{c} h^2$     	& [0.001, 0.99]                  & Cold dark matter density parameter \\
    $100\,\theta_{s}$  				& [0.5, 10]                      & $100 \times$ the angular scale of the sound horizon at decoupling  \\
    $\tau_\mathrm{rei}$ 			& [0.01, 0.8]      				 & Optical depth due to reionization \\
    $n_s$            				& [0.8, 1.2]                     & Scalar spectral index \\
    $\ln(10^{10}A_s)$  				& [1.61, 3.91]                   & Log power of the scalar power spectrum amplitude \\
    $m_\nu$			  				& [0, 1.667]					 & Mass of one of the three degenerate neutrinos\\
	$\kappa_\nu$					& [-1, 1]						 & Statistical parameter\\
    \bottomrule
    \end{tabular}
	\caption{The variable parameters and their prior ranges adpoted in the analysis. }
	\label{tab:priors}
}
\end{table}

\section{Results and Discussions}
\label{Results and Discussions}
In this section, we present the results of our  fitting. 
The constraints on neutrino statistical property, neutrino mass and $H_0$ will be discussed.

The main fitting results are shown in Fig.~\ref{fig:mainbody_figure}. 
 The curves in the top panel of each column present the 1D marginalized densities and the shaded areas are 2D comparison plots showing the correlation between the corresponding two parameters. 
Dark-shaded areas denote the 1$\sigma$ confidence intervals, while light-shaded areas indicate the 2$\sigma$ confidence intervals.

The grey curves and grey shaded areas indicate the fitting results for \textbf{CMB} and the red ones indicate results for \textbf{CMB+BAO}. 
{The 1D marginalized density of $\kappa_\nu$ shows that purely bosonic neutrino is strongly disfavoured by \textbf{CMB}.
Although more weakly, including DESI BAO distance measurements also disfavors $\kappa_\nu=-1$.
The dashed lines show that purely bosonic neutrino can be excluded at 95\% CL by both \textbf{CMB} and \textbf{CMB+BAO}.}
{As shown in the upper panel of the second column in Fig.~\ref{fig:mainbody_figure}, when BAO distance measurements are included, the fitting would prefer slightly higher $H_0$s. The degeneracy between $H_0$ and $\kappa_\nu$ makes relatively smaller $\kappa_\nu$ preferred. Therefore, the constraint from \textbf{CMB+BAO} is weaker than the one from \textbf{CMB}.}
{Both of the improvements achieved here with respect to  results in \cite{deSalas:2018idd} can be attributed to the more accurate datasets and more stringent constraints on $H_0$.}

In the middle panel of the first column in Fig.~\ref{fig:mainbody_figure},  we can see that
there is a correlation between larger $H_0$ and smaller $\kappa_\nu$ for \textbf{CMB} or  between smaller $H_0$ and larger $\kappa_\nu$, 
which is the degeneracy of the effects of $H_0$ and  $\kappa_\nu$ on CMB as discussed in Section 2.
Therefore, a more stringent constraint on $H_0$ would lead to a more stringent constraint on $\kappa_\nu$.
Indeed,  $\kappa_\nu$ is more constrained to a region with larger values when comparing the 
grey dark-shaded area to the grey light-shaded area. 
The exclusion of purely bosonic neutrinos can be attributed to the stringent constraint on $H_0$ by the {CMB} data used.
When comparing the 1D marginalized density of $H_0$ with the one in previous study (the red curve in the up panel of Fig.6 in \cite{deSalas:2018idd}), we find that the constraint on $H_0$ has been strengthened using the latest {CMB} data. Consequently, $\kappa_\nu=-1$ is excluded in our study while it is not in the previous study (the red curve in Fig.4 in \cite{deSalas:2018idd}).

\begin{figure}[htbp]
\centering
\includegraphics[width=0.65\textwidth]{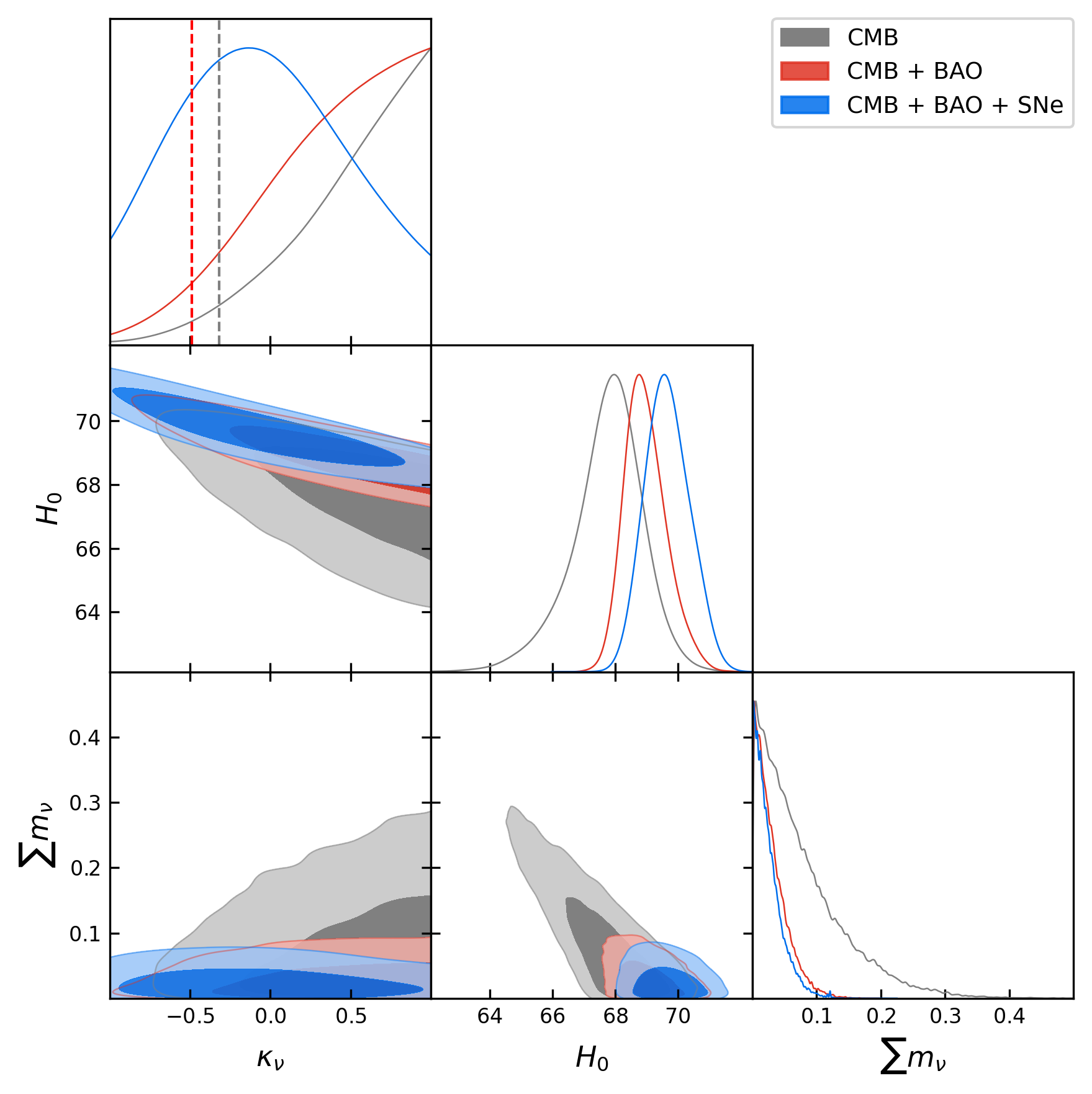}
    \caption{The cosmological constraints for variable-neutrino-statistics model. The grey lines and grey shaded areas in the figure represent the fitting results using only the CMB data. The red ones indicate the fitting results using CMB and BAO data, while the blue ones are results using combined data of CMB, BAO, and local $H_0$ measurement.
Dark-shaded areas denote the 1$\sigma$ confidence intervals, while light-shaded areas indicate the 2$\sigma$ confidence intervals.
{The grey (red) dashed line {in the upper-left panel} is the lower bound on $\kappa_\nu$ at 95\% confidence level for \textbf{CMB} (\textbf{CMB+BAO}).}
}
    \label{fig:mainbody_figure}
\end{figure}

When BAO data is combined with CMB data, 
the degeneracy between $H_0$ and $\kappa_\nu$ remains, as shown by the red area in the 2D plot for $H_0-\kappa_\nu$ in Fig.~\ref{fig:mainbody_figure}. 
Purely bosonic neutrinos are excluded at 2$\sigma$ confidence level along with a more stringent constraint on $H_0$ obtained.

It is important to notice that the correlation between $H_0$ and $\kappa_\nu$ is enhanced for \textbf{CMB+BAO}.
Here, "enhanced correlation" refers to a narrower posterior distribution along the degeneracy direction, reflecting tighter constraints.
This effect may be understood by the stronger constraints on $\sum m_\nu$. 
For \textbf{CMB+BAO}, the constraint on $\sum m_\nu$ is strengthened and a smaller neutrino mass $\sum m_\nu\lesssim 0.1$ eV is preferred,
as shown in the plot of 1D marginalized density of $\sum m_\nu$ in the right column of Fig.~\ref{fig:mainbody_figure}. 
This makes both the $k_\nu-\sum m_\nu$ correlation and the $H_0-\sum m_\nu$ correlation to a narrower region with smaller $\sum m_\nu$,
as can be seen by comparing the grey area with the red area in the lower panel of the first column and the second column in Fig.~\ref{fig:mainbody_figure}.
As discussed before, small neutrino mass, say $\sum m_\nu\lesssim0.1$ eV, would have very small effects on the evolution of the universe.
{So this would break the degeneracy not only between $\sum m_\nu$ and $\kappa_\nu$ but also between $\sum m_\nu$ and $H_0$.}
Consequently, the correlation between $\kappa_\nu$ and $H_0$ is enhanced.

As shown by the blue curve in the 1D plot for $H_0$ in Fig.~\ref{fig:mainbody_figure}, when the local measurement of $H_0$ is also included, the confidence interval for $H_0$  is extended to larger values due to the so-called 'Hubble tension'.
For \textbf{CMB+BAO+SNe}, the constraint on $\sum m_\nu$, the $\kappa_\nu-\sum m_\nu$ correlation and 
$H_0-\sum m_\nu$ correlation remain roughly the same as the ones for \textbf{CMB+BAO} respectively.  
{In contrast, as shown in the 2D plot for $H_0$ and $\kappa_\nu$, the correlation between them has been shifted significantly and the purely B-E neutrinos become possible at 2$\sigma$ confidence level.}
However, due to the presence of the Hubble tension, the \textbf{CMB+BAO+SNe} may not provide robust support for physical interpretation and should therefore be treated with caution.
We present its results only for reference and further discussion.

\begin{figure}[htbp]
\centering
\begin{subfigure}[b]{0.48\textwidth}
\centering
\includegraphics[width=\textwidth]{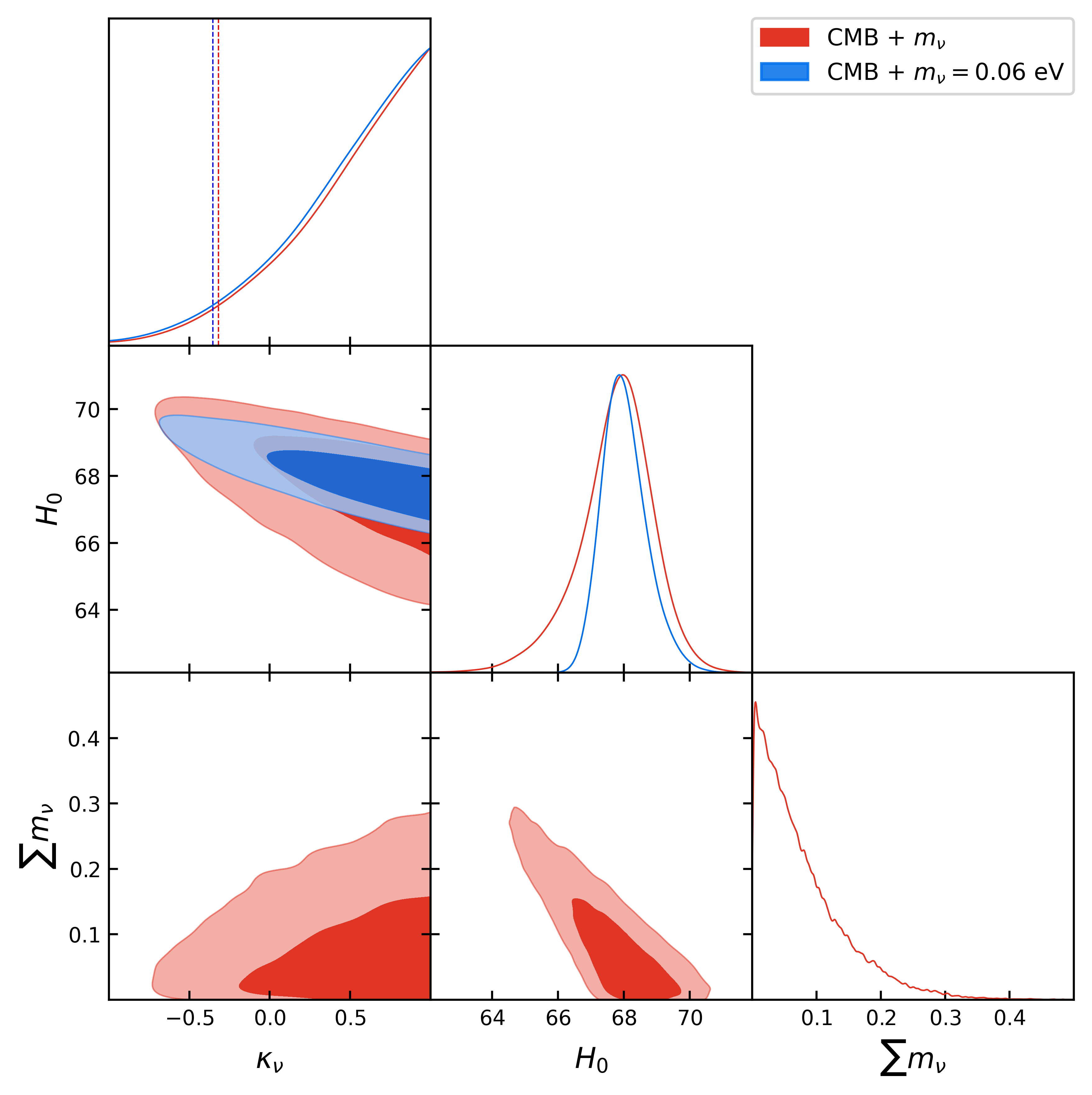}
\caption{}
\label{fig:neutrino_mass_fixed_figuresub1}
\end{subfigure}
\hfill
\begin{subfigure}[b]{0.48\textwidth}
\centering
\includegraphics[width=\textwidth]{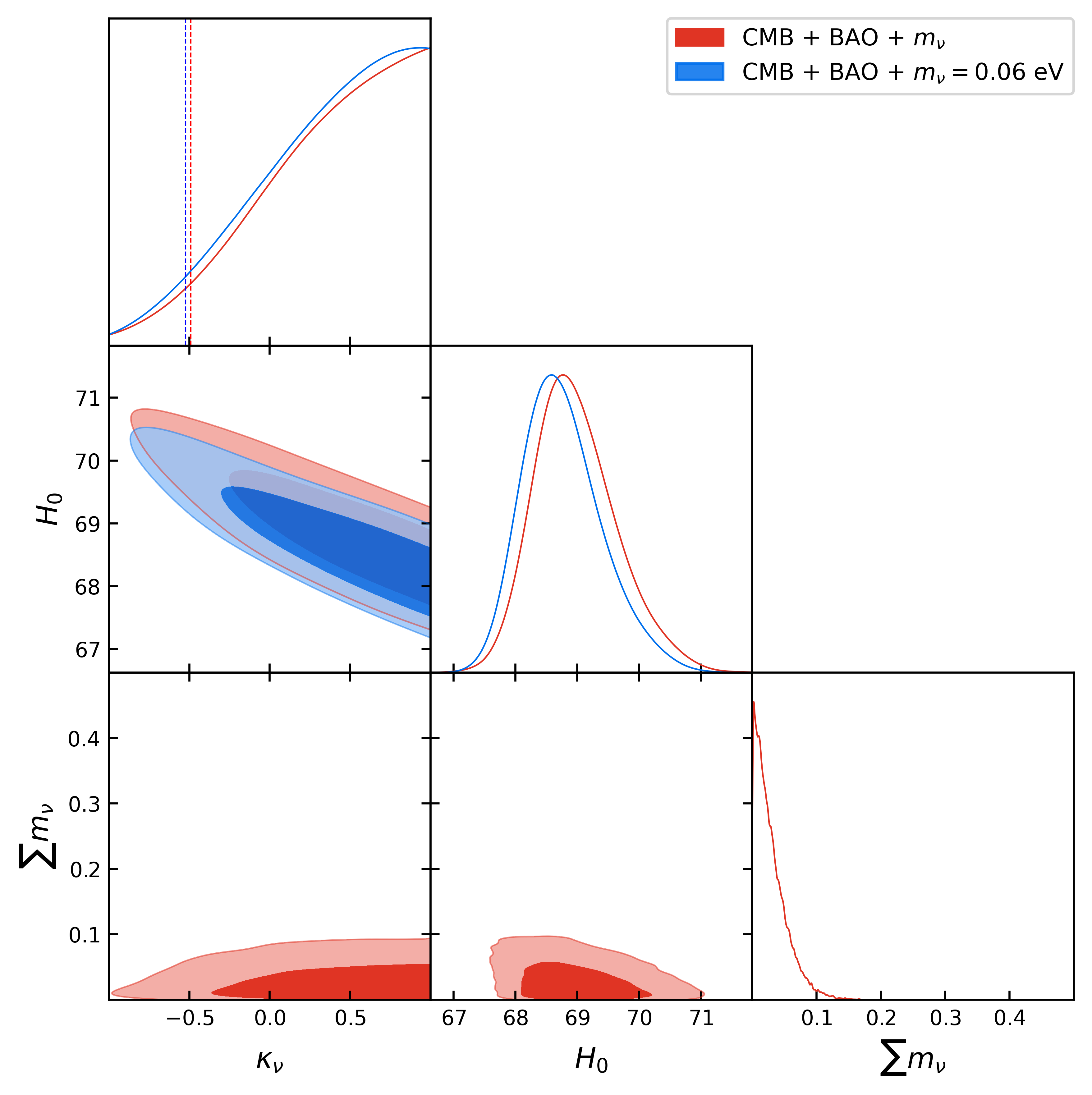}
\caption{}
\label{fig:neutrino_mass_fixed_figuresub2}
\end{subfigure}
\caption{{The cosmological constraints for variable-neutrino-statistics model. The left panel contains the results for \textbf{CMB} only while the right one contains the results for \textbf{CMB+BAO}. The red lines and red shaded areas in the figures are fitting results allowing $\sum m_\nu$ to vary, while the blue ones are for fixed $\sum m_\nu=0.06$ eV.}
The {CMB+$m_\nu$} and {CMB+BAO+$m_\nu$} present the fitting results for the $\sum m_\nu$ variable model using CMB data and CMB+BAO data respectively, while {CMB+$m_\nu=0.06$ eV} and {CMB+BAO+$m_\nu=0.06$ eV} present the corresponding results for the model with $\sum m_\nu$ fixed at 0.06 eV. }
\label{fig:neutrino_mass_fixed_figure}
\end{figure}

To further investigate the impact of neutrino mass on $\kappa_\nu$, 
we perform a fit with $\sum m_\nu=0.06\ \mathrm{eV}$ fixed, and compare it to the case with variable neutrino mass.
The comparison is made in Fig.~\ref{fig:neutrino_mass_fixed_figure}.
The results obtained for the {CMB} dataset are shown in Fig.~\ref{fig:neutrino_mass_fixed_figure}(a) 
and the results obtained for the {CMB+BAO} dataset are shown in Fig.~\ref{fig:neutrino_mass_fixed_figure}(b).
The red curves and shaded areas denote the fitting results of variable neutrino mass model; the blue ones denote the results of $\sum m_\nu = 0.06$ eV fixed model.

Comparing the blue area and the red area in the middle panel of the first column in Fig.~\ref{fig:neutrino_mass_fixed_figure}(a) we find that the correlation between $H_0$ and $\kappa_\nu$ is enhanced once $\sum m_\nu$ is fixed.
As discussed for Fig.~\ref{fig:Hubble_normalized_mass_vary} and Fig.~\ref{fig:rs_da_H0_theta}, 
large $\sum m_\nu$ would have considerable effects on $H(a)$ and $\theta_s$.
So there exists a degeneracy between $\sum m_\nu$ and $H_0$ if  $\sum m_\nu$ is large enough.
Once $\sum m_\nu$ is fixed, this degeneracy would be broken.
Therefore, the correlation between $\kappa_\nu$ and $H_0$ is enhanced.

However, this effect becomes less pronounced after incorporating the BAO data. 
As seen in the right panel of Fig.~\ref{fig:neutrino_mass_fixed_figure}, 
the 2D comparison plots for $\kappa-H_0$ degeneracies 
in the two analyses—one with a fixed neutrino mass and the other allowing the mass to vary—are quite similar.
The reason is that the DESI BAO measurement favors smaller neutrino mass, as can be seen in the plot of 1D marginalized density of $\sum m_\nu$ 
in Fig.~\ref{fig:mainbody_figure}.
As discussed earlier, low-mass neutrinos become non-relativistic at very late times, and their impact on $\theta_s$ is negligible.  
As a result, whether $\sum m_\nu$ slightly varies in a small mass regime or not would not significantly affect $\theta_s$. 
That is why the degeneracies in the two analyses—one with a fixed neutrino mass and the other allowing the mass to vary—are quite similar.

\begin{table}[h!]
\centering
% 调整行高
\renewcommand{\arraystretch}{1.5} % 数值越大，行高越大
\begin{tabular}{|c|c|c|c|c|c|}
\hline
                       & $\kappa_\nu$ mean  & $\kappa_\nu$ lower bound  &$\sum m_\nu$ mean   &$\sum m_\nu$ upper bound\\ \hline
CMB+$m_\nu$            & 0.479       & -0.317      &0.0772         &0.2114\\ \hline
CMB+BAO+$m_\nu$        & 0.354       & -0.489      &0.0294         &0.0784\\ \hline
CMB+BAO+SNe+$m_\nu$           & -0.051       & -1          &0.0253         &0.0682         \\ \hline
CMB+$m_\nu=0.06$ eV     & 0.465       & -0.352      &/              &/\\ \hline
CMB+BAO+$m_\nu=0.06$ eV & 0.336       & -0.523      &/              &/         \\ \hline
\end{tabular}
\caption{Fitting results for the cosmological parameters \( \kappa_\nu \) and \( \sum m_\nu \) .   The second and third column show the mean value of $\kappa_\nu$ and the lower bound on \( \kappa_\nu \) at 95\% confidence level.
 The last two columns provide the mean value of  \(\sum m_\nu \) and its upper bound at 95\% confidence level.
} 
\label{tab:mean_value_and_bound}
\end{table}

\begin{table}[h!]
\centering
% 调整行高
\renewcommand{\arraystretch}{1.5} % 数值越大，行高越大
\begin{tabular}{|c|c|c|c|c|c|c|c|}
\hline
                       &$\kappa_\nu$&$H_0 $&$\sum m_\nu$[eV]&$\chi^2$&$\chi^2_\mathrm{CMB}$&$\chi^2_\mathrm{BAO}$   &$\chi^2_\mathrm{SNe}$\\ \hline
CMB+$m_\nu$            & 0.60       & 67.92      & 0.082      & 2796.13   &2796.13  &/        &/ \\ \hline
CMB+BAO+$m_\nu$        & 0.93       & 68.36      & 0.010      & 2809.97   &2795.44  &14.53    &/ \\ \hline
CMB+BAO+SNe+$m_\nu$    & -0.04      & 69.40      & 0.034      & 3852.08   &2799.22    &13.50  &1039.35 \\ \hline
CMB+$m_\nu=0.06$ eV     & 0.41       & 67.87      & 0.06       & 2796.14   &2796.14  &/       &/ \\ \hline
CMB+BAO+$m_\nu=0.06$ eV & 0.98       & 68.38      & 0.06       & 2812.85   &2798.80  &14.05   &/ \\ \hline
\end{tabular}
\caption{Best-fit parameter values and corresponding \(\chi^2\) values for different datasets. The columns include the values of \(\kappa_\nu\), \(H_0\), the sum of neutrino mass \(\sum m_\nu\), and the \(\chi^2\) values for the total dataset, CMB, and BAO data.
The reported $\chi^2$ values are obtained from the MCMC chains.
The Hubble constant $H_0$ is expressed in units of km\,s$^{-1}$\,Mpc$^{-1}$.}
\label{tab:The best fit values}
\end{table}

The constraints on the neutrino statistical parameter $\kappa_\nu$  and neutrino mass $\sum m_\nu$ from different datasets are summarized in Table ~\ref{tab:mean_value_and_bound}. All the fitting results rule out the possibility of purely bosonic neutrinos at 95\% confidence level except the one using {CMB+BAO+SNe} dataset, whose fitting quality may be poor due to the Hubble tension.
Neutrino with mixed statistics remains as a viable option with $-0.5\lesssim\kappa_\nu<1$ at $2\sigma$ confidence interval.
Comparing the upper bound of $\sum m_\nu$ given by {CMB+BAO+$m_\nu$} fitting to the results in Table 4 of \cite{DESI:2024mwx}, it can be found that introducing mixed statistics for neutrinos causes the DESI data to favor a slightly larger neutrino mass.
The upper bound increases from 0.072 eV to 0.078 eV. This helps reduce the potential neutrino mass tension between cosmological observations and neutrino oscillation experiments, but the effect remains minimal.

Table~\ref{tab:The best fit values}  presents the results of best-fit parameters for each dataset, along with their corresponding $\chi^2$  values. 
{According to the table, the results from CMB+BAO+$m_\nu$ and CMB+BAO+~$m_\nu=0.06$ eV combinations indicate that fermionic statistics is preferred by \textbf{CMB+BAO}. 
In contrast, the results of CMB+$m_\nu$ and CMB+$m_\nu=0.06$ eV suggest a preference for a mixed distribution by \textbf{CMB}.
}
The results of CMB+BAO+SNe+$m_\nu$ suggest a stronger preference for the mixed statistics neutrino model.
However, due to the presence of the Hubble tension in this dataset combination, this result should be viewed with caution and considered for reference only.

{All fittings in this study have been performed under the assumption that neutrinos have degenerate masses.
However, neutrino mass hierarchies might have effects on the evolution of the universe.  
To clarify this point, we have performed a simplified fitting with two $0.01$ eV neutrinos and one $0.04$ eV neutrino fixed using {CMB+BAO} data.
The fitting result is nearly the same as that of {CMB+BAO+$m_\nu$=0.06 eV}.
This agrees with a recent analysis \cite{herold2024revisitingimpactneutrinomass} .
Thus, the degenerate-neutrino-mass approximation in our study is a reasonable assumption.
}

\section{Conclusions}
\label{Conclusions}

In this work, we have investigated effects of statistical property and mass of neutrino in cosmology. We focus on how different datasets, including CMB and BAO measurements, can constrain the sum of the three degenerate neutrino masses $\sum m_\nu$, the neutrino statistical parameter $\kappa_\nu$ and the present Hubble rate $H_0$.

In Section~\ref{Neutrino Statistics and mass in Cosmology}, we discuss the effects of $\sum m_\nu$, $\kappa_\nu$ and $H_0$ on the Hubble rate $H(a)$, the sound horizon $r_s(\eta)$, the angular diameter $D_A(\eta)$ at recombination and the peak scale parameter $\theta_s(\eta)$.
We show that increasing any of the three parameters would have similar influences on $\theta_s(\eta)$ so we expect a degeneracy among them.

We fit the variable-neutrino-statistics model with {CMB}, {CMB+BAO} and {CMB+BAO+SNe} datasets respectively.
{We find that purely bosonic neutrinos have been excluded by the {CMB} or {CMB+BAO} datasets at $2\sigma$.}
This is our first key conclusion.

Based on the mean values and the best fit results of $\kappa_\nu$, 
{we conclude that CMB+BAO dataset favors neutrinos following Fermi-Dirac statistics, while neutrinos with mixed statistics remain a possibility especially when only CMB data are included in the fit.}
This forms our second conclusion.

Looking ahead, the release of more precise cosmological data from upcoming experiments like {CMB-S4} \cite{schiappucci2024constrainingcosmologicalparametersusing} and {Euclid} \cite{euclidcollaboration2024euclidiovervieweuclid} is expected to strengthen the constraints on $H_0$, potentially enabling more robust limits on mixed-statistics neutrinos. Furthermore, should the Hubble tension be fully resolved in the future, it could lead to a more accurate understanding of neutrino statistics and refine our models of cosmological evolution.

\bigskip
\section*{Note added}
\label{sec:Note added}
After the submission of this article, the DESI Collaboration made their second data release (DESI DR2) \cite{desicollaboration2025desidr2resultsii}.
We have combined the DESI DR2 data with CMB data and repeated the fitting \textbf{CMB+BAO} defined in Section~\ref{Data} of the main text.
We find no significant differences compared to the previous results, as can be seen in Fig. \ref{fig:comparison} and Fig.~\ref{fig:CMBBAO006_comparison}.
Our first conclusion remains valid, while the second one is weakened slightly because the best fit value of $\kappa_\nu$ moves to a slightly smaller value, as seen in the third line in Table~\ref{tab:comparison}.

\begin{figure}[htbp]
\centering
\includegraphics[width=0.65\textwidth]{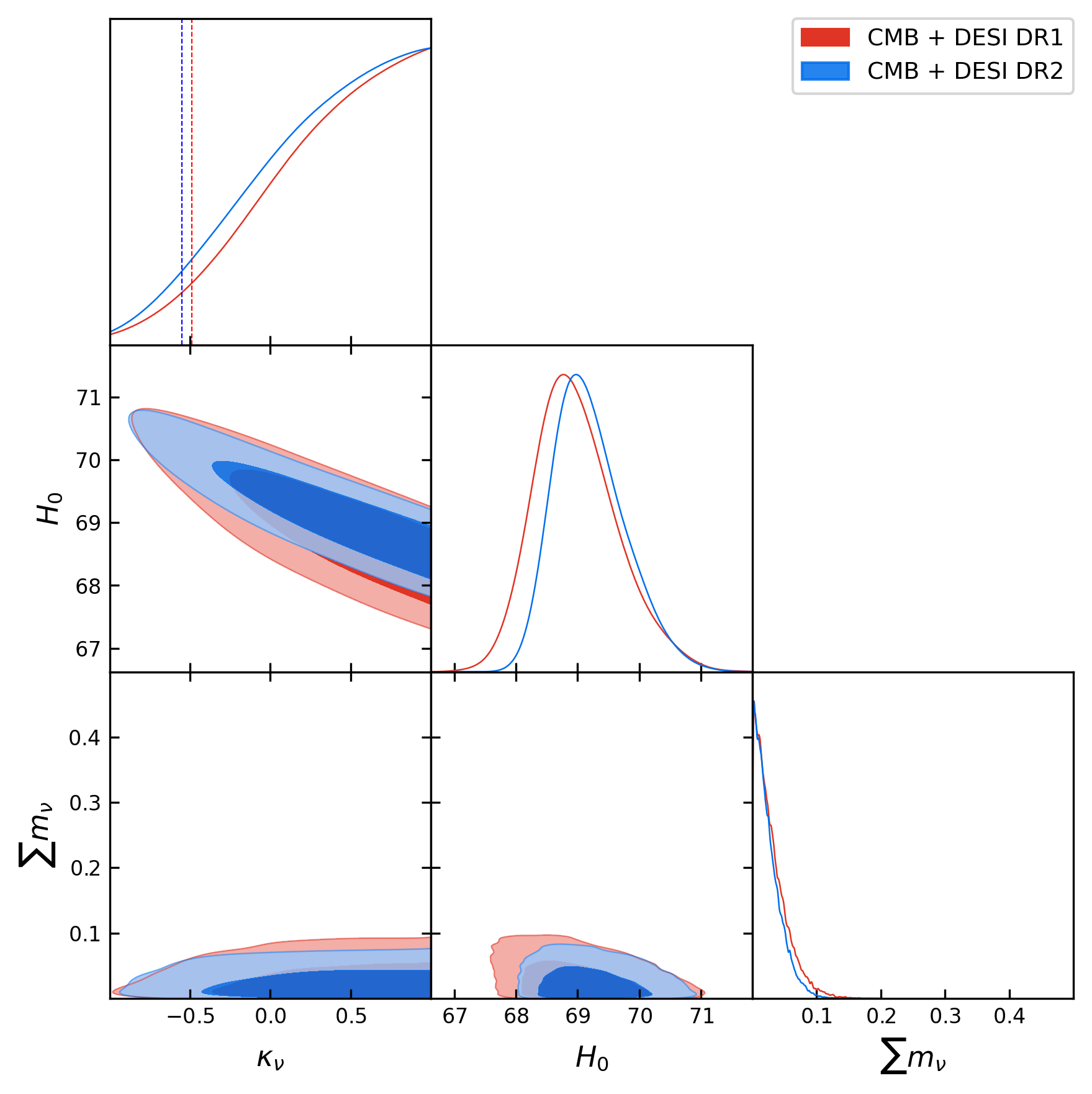}
    \caption{
Comparison of the results for \textbf{CMB+BAO} using the DESI DR1 data and the newly released DESI DR2 data.
Dark-shaded areas denote the 1$\sigma$ confidence intervals, while light-shaded areas indicate the 2$\sigma$ confidence intervals.
The red (blue) dashed line {in the upper-left panel} is the lower bound on $\kappa_\nu$ at 95\% confidence level for \text{CMB + DESI DR1} (\text{CMB + DESI DR2}).
}
    \label{fig:comparison}
\end{figure}

\begin{figure}[htbp]
\centering
\includegraphics[width=0.65\textwidth]{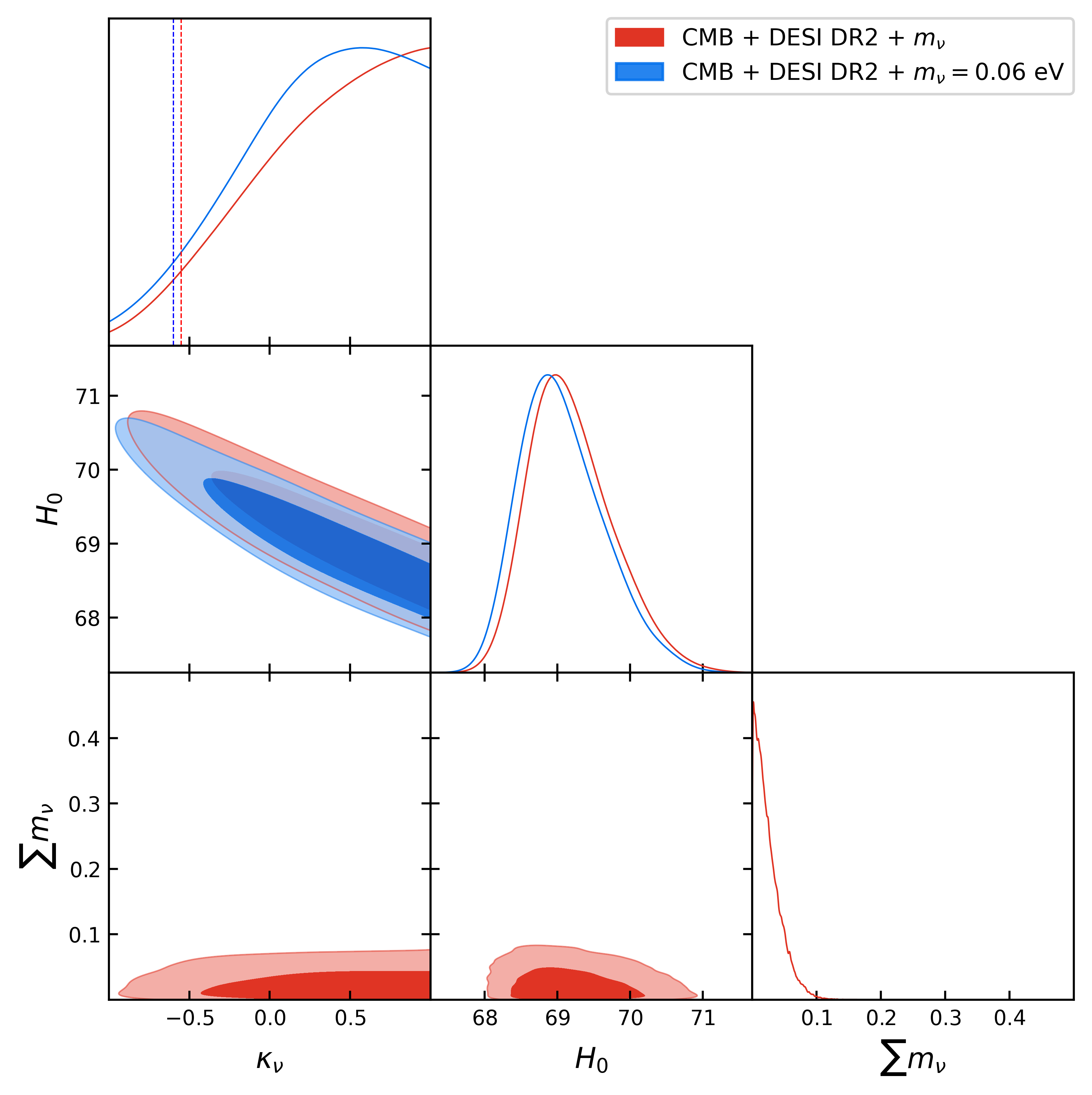}
    \caption{
The cosmological constraints for variable-neutrino-statistics model. 
This figure is identical to Fig.~\ref{fig:neutrino_mass_fixed_figure} in the main text, except that the BAO data have been replaced with DESI DR2.
}
    \label{fig:CMBBAO006_comparison}
\end{figure}

\begin{table}[h!]
\centering
% 调整行高
\renewcommand{\arraystretch}{1.5} % 数值越大，行高越大
\begin{tabular}{|c|c|c|c|c|c|c|}
\hline
                     & $\kappa_\nu$              	 &$\sum m_\nu$ [eV]		        &$H_0$ [km $\mathrm{s}^{-1}$ $\mathrm{Mpc}^{-1}$]
					 &$\chi^2$						 &$\chi^2_{\mathrm{CMB}}$		&$\chi^2_{\mathrm{BAO}}$ \\ \hline
CMB+DESI DR1         & 0.93						     &0.010 						&68.36  			&2809.97 &2795.44 &14.53 \\ \hline
CMB+DESI DR2         & 0.76      					 &0.018 						&68.64              &2807.00 &2795.34 &11.66 \\ \hline
CMB+DESI DR1+$m_\nu = 0.06$ eV
					 & 0.98      					 &/ 						    &68.38              &2812.85 &2798.80 &14.05 \\ \hline
CMB+DESI DR2+$m_\nu = 0.06$ eV
					 & 0.45      					 &/    						    &68.94              &2812.45 &2801.31 &11.14 \\ \hline
\end{tabular}
\caption{Comparison of the best-fit values of \( \kappa_\nu \), \( \sum m_\nu \) and \(H_0\).
} 
\label{tab:comparison}
\end{table}

\bigskip
\section*{Acknowledgements}
\label{sec:acknowledgements}
W. Liao is supported by National Natural Science Foundation of China under the grant No. 11875130.

The authors are grateful to the anonymous referee for the careful reading of the manuscript and many insightful suggestions.

\end{document}